\documentclass[twocolumn,twocolappendix]{aastex63}

\newcounter{daggerfootnote}

\newcommand{\srcname}{GLIMPSE-17775}

\newcommand{\jwst}{\textit{JWST}}
\newcommand{\hst}{\textit{HST}}

\definecolor{mycol}{rgb}{0,0,1}

\received{n/a}
\revised{n/a}
\accepted{n/a}

\submitjournal{\apj}

\shorttitle{Pumping Iron}
\shortauthors{Kokorev et al.}

\usepackage{xcolor}
\usepackage{longtable}
\usepackage{lipsum}
\usepackage{amsmath}
\usepackage[normalem]{ulem}
\usepackage[para]{threeparttable}
\usepackage{enumitem}
\usepackage[encapsulated]{CJK}

\begin{document}

\title{The Deepest GLIMPSE of a Dense Gas Cocoon Enshrouding a Little Red Dot}

\correspondingauthor{Vasily Kokorev}
\email{vkokorev@utexas.edu}

\author[0000-0002-5588-9156]{Vasily Kokorev}
\affiliation{Department of Astronomy, The University of Texas at Austin, Austin, TX 78712, USA}
\affiliation{Cosmic Frontier Center, The University of Texas at Austin, Austin, TX 78712, USA} 

\author[0000-0002-0302-2577]{John Chisholm}
\affiliation{Department of Astronomy, The University of Texas at Austin, Austin, TX 78712, USA}
\affiliation{Cosmic Frontier Center, The University of Texas at Austin, Austin, TX 78712, USA} 

\author[0000-0003-3997-5705]{Rohan P.~Naidu}
\affiliation{MIT Kavli Institute for Astrophysics and Space Research, 70 Vassar Street, Cambridge, MA 02139, USA}

\author[0000-0001-7201-5066]{Seiji Fujimoto}
\affiliation{David A. Dunlap Department of Astronomy and Astrophysics, University of Toronto, 50 St. George Street, Toronto, Ontario, M5S 3H4, Canada}
\affiliation{Dunlap Institute for Astronomy and Astrophysics, 50 St. George Street, Toronto, Ontario, M5S 3H4, Canada}

\author[0000-0002-7570-0824]{Hakim Atek}
\affiliation{Institut d'Astrophysique de Paris, CNRS, Sorbonne Universit\'e, 98bis Boulevard Arago, 75014, Paris, France}

\author[0000-0003-2680-005X]{Gabriel Brammer}
\affiliation{Cosmic Dawn Center (DAWN), Niels Bohr Institute, University of Copenhagen, Jagtvej 128, K{\o}benhavn N, DK-2200, Denmark}

\author[0000-0001-8519-1130]{Steven L. Finkelstein}
\affiliation{Department of Astronomy, The University of Texas at Austin, Austin, TX 78712, USA}
\affiliation{Cosmic Frontier Center, The University of Texas at Austin, Austin, TX 78712, USA} 

\author[0000-0003-3596-8794]{Hollis B. Akins}
\affiliation{Department of Astronomy, The University of Texas at Austin, Austin, TX 78712, USA}
\affiliation{Cosmic Frontier Center, The University of Texas at Austin, Austin, TX 78712, USA} 

\author[0000-0002-4153-053X]{Danielle A. Berg}
\affiliation{Department of Astronomy, The University of Texas at Austin, 2515 Speedway, Stop C1400, Austin, TX 78712, USA}
\affiliation{Cosmic Frontier Center, The University of Texas at Austin, Austin, TX 78712, USA} 

\author[0000-0001-6278-032X]{Lukas J. Furtak}
\affiliation{Department of Astronomy, The University of Texas at Austin, Austin, TX 78712, USA}
\affiliation{Cosmic Frontier Center, The University of Texas at Austin, Austin, TX 78712, USA} 

\author[0000-0001-7232-5355]{Qinyue Fei}
\affiliation{David A. Dunlap Department of Astronomy and Astrophysics, University of Toronto, 50 St. George Street, Toronto, Ontario, M5S 3H4, Canada}

\author[0000-0003-4512-8705]{Tiger Yu-Yang Hsiao}
\affiliation{Department of Astronomy, The University of Texas at Austin, Austin, TX 78712, USA}
\affiliation{Cosmic Frontier Center, The University of Texas at Austin, Austin, TX 78712, USA} 

\author[0000-0002-2057-5376]{Ivo Labb\'e}
\affiliation{Centre for Astrophysics and Supercomputing, Swinburne University of Technology, Melbourne, VIC 3122, Australia}

\author[0000-0003-2871-127X]{Jorryt Matthee} 
\affiliation{Institute of Science and Technology Austria (ISTA), Am Campus 1, 3400 Klosterneuburg, Austria}

\author[0000-0002-8984-0465]{Julian B.~Mu\~noz}
\affiliation{Department of Astronomy, The University of Texas at Austin, Austin, TX 78712, USA}
\affiliation{Cosmic Frontier Center, The University of Texas at Austin, Austin, TX 78712, USA} 

\author[0000-0001-5851-6649]{Pascal A. Oesch}
\affiliation{D\'epartement d'Astronomie, Universit\'e de Gen\`eve, Chemin Pegasi 51, 1290 Versoix, Switzerland}
\affiliation{Cosmic Dawn Center (DAWN), Niels Bohr Institute, University of Copenhagen, Jagtvej 128, K{\o}benhavn N, DK-2200, Denmark}

\author[0000-0002-9651-5716]{Richard Pan}
\affiliation{Department of Physics \& Astronomy, Tufts University, MA 02155, USA}

\author[0000-0002-5104-8245]{Pierluigi Rinaldi}
\affiliation{Steward Observatory, University of Arizona, 933 North Cherry Avenue, Tucson, AZ 85721, USA}

\author[0000-0001-8419-3062]{Alberto Saldana-Lopez}
\affiliation{Department of Astronomy, The Oskar Klein Centre, Stockholm University, AlbaNova, SE-10691 Stockholm, Sweden}

\author[0000-0001-7144-7182]{Daniel Schaerer}
\affiliation{Observatoire de Genève, Université de Genève, Chemin Pegasi 51, 1290 Versoix, Switzerland}
\affiliation{CNRS, IRAP, 14 Avenue E. Belin, 31400 Toulouse, France}

\author[0000-0002-3216-1322]{Marta Volonteri}
\affiliation{Institut d'Astrophysique de Paris, CNRS, Sorbonne Universit\'e, 98bis Boulevard Arago, 75014, Paris, France}

\author[0000-0002-0350-4488]{Adi Zitrin}
\affiliation{Department of Physics, Ben-Gurion University of the Negev, P.O. Box 653, Be'er-Sheva 84105, Israel}

\begin{abstract}
The detection of strong Balmer breaks and absorption features in Little Red Dots (LRDs) suggests they host AGN embedded within dense gas envelopes, potentially powered by super-Eddington accretion. We present GLIMPSE-17775, a luminous ($L_{\rm bol}\sim10^{45}$ erg s$^{-1}$) LRD at $z=3.501$ behind Abell S1063 ($\mu\sim2$), observed with deep \textit{JWST}/NIRCam and a $\sim$20 hr (80 hr de-lensed) NIRSpec/G395M spectrum. The data reveal 40+ emission and absorption features, including a rich forest of low-ionization Fe\,\textsc{ii} lines and numerous broad hydrogen recombination transitions. We use this depth to test the dense-gas interpretation through five independent diagnostics. Nearly all permitted lines show exponential wings with consistent FWHM, the signature of Thomson scattering requiring $n_e\gtrsim10^8$ cm$^{-3}$. Adopting this width yields $M_{\rm BH}\sim10^{6.7}M_\odot$, a factor of ten lower than Gaussian fits, and $\lambda_{\rm Edd}\sim1.8$. Additional diagnostics support the same picture: a pronounced Balmer break ($f_{\nu,4050}/f_{\nu,3670}=2.0\pm0.1$), enhanced He\,\textsc{i} $\lambda7065$ and $\lambda10830$ with P-Cygni absorption, Bowen-fluorescent O\,\textsc{i} $\lambda8446$–$\lambda11290$ emission requiring Ly$\beta$ pumping, and 16 Fe\,\textsc{ii} lines matching fluorescence models. These features indicate a dense ($n\sim10^8$ cm$^{-3}$), partially ionized cocoon where scattering and fluorescence dominate line formation, providing strong evidence that at least some LRDs are powered by super-Eddington black-hole growth in the early Universe.
\end{abstract}


\section{Introduction}
\label{sec:intro}
One of the most enticing puzzles brought upon by the launch of the \textit{James Webb Space Telescope} (\textit{JWST}) has been the discovery of red, compact objects called ``Little Red Dots" \citep[LRDs;][]{matthee23}. Previously invisible to the \textit{Hubble Space Telescope} (\textit{HST}) due to their extreme faintness in the optical and lack of near-infrared (NIR) coverage, LRDs have emerged in abundance \citep{kokorev24,kocevski24,akins24,barro24} thanks to unprecedented NIR sensitivity of \textit{JWST}. 

Their unusual properties—such as compact morphologies in rest-optical and distinctive “v-shaped” spectral energy distributions (SEDs)—make LRDs easy to identify in \textit{JWST} fields, however this is where the simplicity ends. Explaining LRDs as either evolved or dusty compact galaxies proves difficult: the former scenario requires massive stellar populations that strain $\Lambda$CDM predictions \citep{labbe23_nat,boylan-kolchin23}, while the latter implies significant dust emission—yet none is observed \citep{labbe23,casey24_lrd,akins24,leung24}. 

Over time, the accumulating detections of broad Balmer series emission lines \citep[][just to name a few]{kocevski23,kocevski24,matthee23}, often accompanied by signatures of extreme ionization \citep[e.g.,][]{kokorev23c}, have begun to clarify the physical origin of LRDs, pointing increasingly toward AGN as the underlying power source. In parallel, the much-needed advent of NIRSpec MSA programs based on red targets selected from JWST imaging \citep[e.g.,][]{rubies24} has provided critical confirmation: nearly all point sources exhibiting “v-shaped” spectral energy distributions (SEDs) reveal broad emission lines upon spectroscopic follow-up \citep{hviding25}, solidifying their AGN interpretation.

The nature of the spectral inflection point in LRDs, typically located near $\sim3600$ \AA, has also undergone significant revision. Initially interpreted as a stellar Balmer break \citep{labbe23_nat}, this feature implied implausibly high stellar masses ($M_*$) far too early in cosmic history \citep{boylan-kolchin23,sabti24}. A second hypothesis invoked differential dust attenuation and host galaxy contamination in the rest-UV to explain the sharp discontinuity \citep{volonteri25}. However, this explanation was soon ruled out by \citet{setton24,ma25}, who demonstrated that the break generally lies around the Balmer limit, albeit with a fundamentally different origin than initially proposed.

More recently, a series of theoretical and observational works have converged on a new picture: that LRDs may host accreting black holes enshrouded in exceptionally dense, partially ionized gas cocoons—the so-called Black Hole Star (BH$^*$) scenario \citep{inayoshi25,naidu25_bh*,taylor25,degraaff25_cliff}. In this framework, the steep Balmer-limit break, blueshifted absorption, and suppressed X-ray and radio emission \citep[e.g][]{naidu25_bh*,inayoshi25} all arise naturally from a radiation-dominated, optically thick environment surrounding a rapidly growing black hole, while the UV emission and forbidden optical lines (e.g. [O\,\textsc{iii}]$\, \lambda \lambda 4959, 5007$) potentially originate from the extended host.  The dense-gas interpretation also offers a natural explanation for the P-Cygni-like Balmer and helium profiles frequently seen in high-S/N spectra \citep{matthee23}, the non-Gaussian, exponentially-winged line shapes predicted by radiative-transfer models of scattering in ionized gas \citep{rusakov25,chang25} and Lyman$\alpha$ - like resonantly scattered shapes \citep{chang25,naidu25_bh*}. Intriguingly, if electron and resonant scattering indeed dominate the broad-line widths, then the true virial velocities (traced by intrinsically narrower Gaussian core)—and hence black-hole masses—could be lower by an order of magnitude, alleviating the apparent tension between LRD black-hole masses and their compact host galaxies \citep{rusakov25}.

So far, however, the BH$^*$ / dense-gas interpretation of LRDs has relied largely on the Balmer break, occasional absorption features \citep{lin25}, and, to a limited extent, the shapes of broad emission lines \citep{rusakov25}. While these features are consistent with the presence of dense, partially ionized gas, they stop short of providing direct spectroscopic confirmation of the physical conditions expected by the BH$^*$ scenario. What has been missing is an unambiguous demonstration—through emission-line physics—that LRDs indeed host dense, optically thick cocoons surrounding rapidly accreting black holes. Such evidence would directly tie the observed line formation, excitation, and radiative transfer to dense, stratified envelopes of gas, surrounding the central engine.

In this work, we present precisely such a case: an exceptionally deep $\sim$20 hr, equivalent to 80 hr without lensing magnification, \textit{JWST}/NIRSpec G395M spectrum of a luminous LRD at $z=3.50102$. The target, \srcname , lies in a highly magnified region of the massive galaxy cluster Abell S1063 (AS1063 hereafter) and benefits from the combined power of \textit{JWST}/NIRCam imaging from the GLIMPSE GO program (PID 3293; PIs H. Atek \& J. Chisholm) and recent NIRSpec spectroscopy from the GLIMPSE-D DDT program (PID 9223; PIs S. Fujimoto \& R. Naidu). This synergy of ultra-deep spectroscopy and strong gravitational lensing enables an unprecedented view of \srcname, revealing rest-frame optical and near-infrared features at a level unseen before in any LRD.
We detect over forty emission and absorption features—sixteen of them Fe \textsc{ii} transitions forming a dense “iron forest.”
The remarkable richness of this spectrum makes \srcname\ a uniquely powerful laboratory for dissecting the dense, radiation-dominated environments that accompany early black-hole growth.

This paper is organized as follows. In \autoref{sec:obs_data} we present the GLIMPSE-D NIRSpec dataset alongside the photometric data sets used in this study. In \autoref{sec:emline_analysis} we 
calculate the spectroscopic redshift, describe the identification of prominent emission/absorption features and describe bespoke fitting of various line complexes. \autoref{sec:data_analysis} describes the measurement  of the source morphology, dust attenuation and black hole masses. In \autoref{sec:emline_props} we comment on the various properties of the emission features in the our target and finally discuss our findings in \autoref{sec:concl}. 

Throughout this work we assume a flat $\Lambda$CDM cosmology with $\Omega_{\mathrm{m},0}=0.3$, $\Omega_{\mathrm{\Lambda},0}=0.7$ and H$_0=70$ km s$^{-1}$ Mpc$^{-1}$, and a \citet{chabrier} IMF between $0.1-100$ $M_{\odot}$. All magnitudes are expressed in the AB system \citep{oke74}.

\begin{figure*}[th]
\centering
\qquad\includegraphics[width=.9\textwidth]{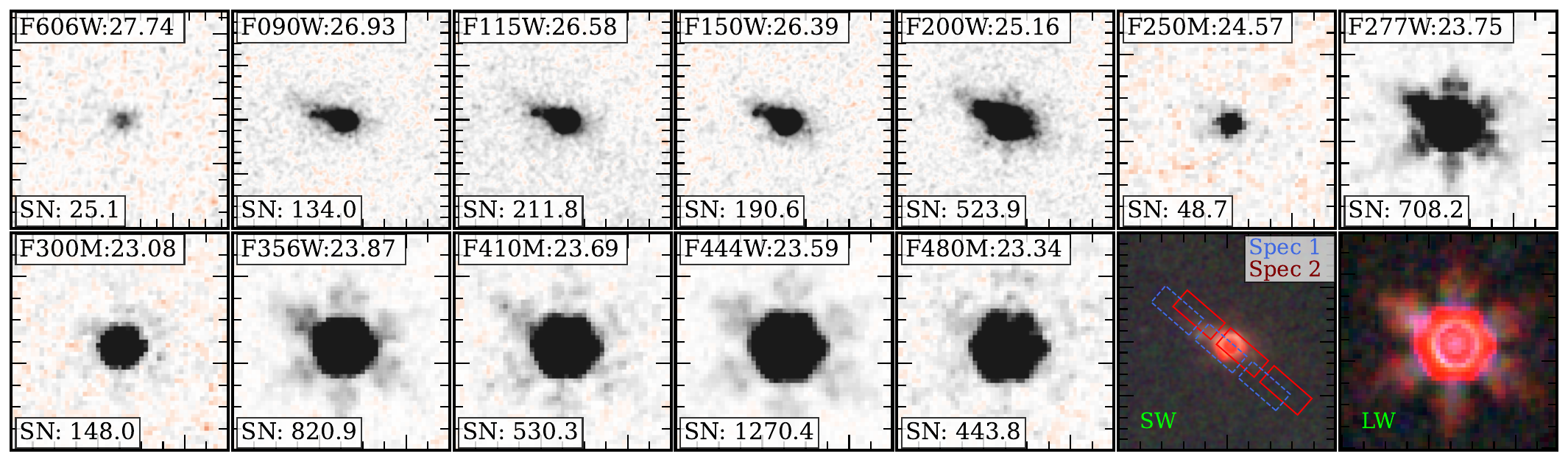}\\[-1.3mm]
\includegraphics[width=.99\textwidth]{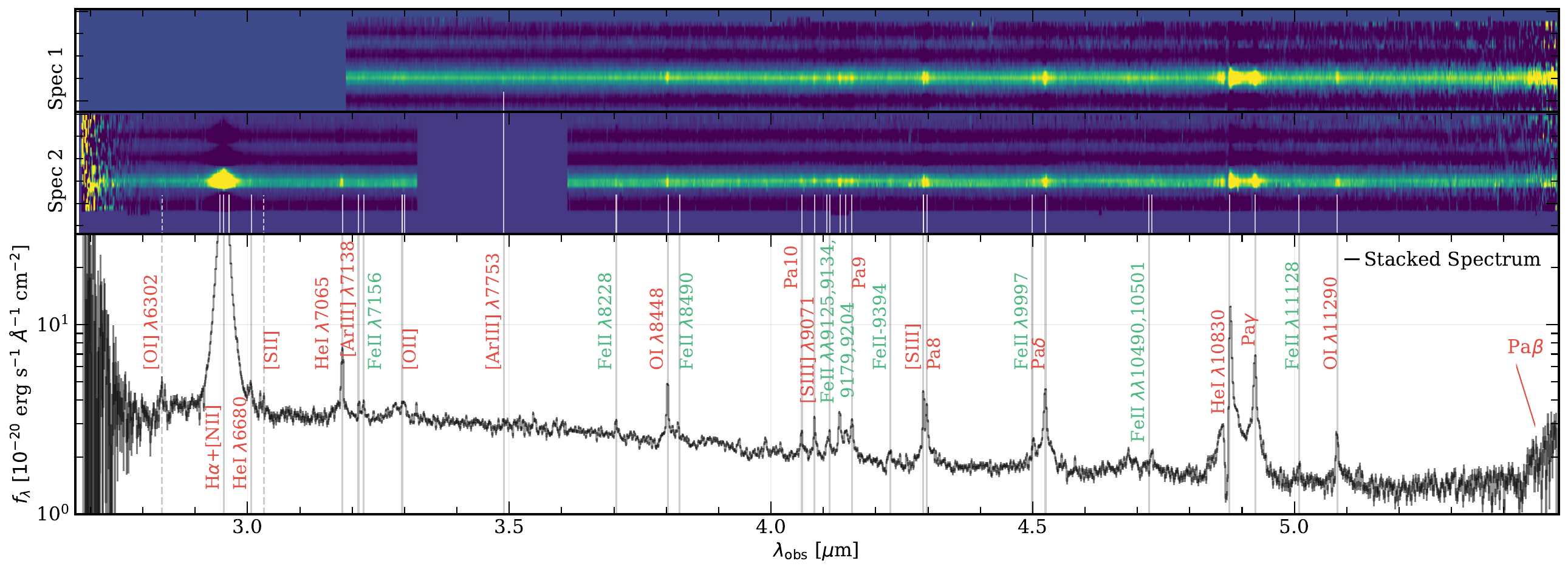}
\caption{\textbf{Top:} \textit{JWST}/NIRCam  and \textit{HST} 2\farcs{0} stamps and the RGB short (SW) and long wavelength (LW) color images comprised of the F115W, F150W, F200W and F277W, F356W, and F444W bands, respectively. MSA shutters for both configurations covering \srcname\ are shown in blue and red, respectively. The source morphology is resolved and extended up to $\sim2\mu$m, and then appears to transition to a more PSF-dominated and compact shape, echoing a growing sample of LRDs with extended rest-UV morphology \citep{matthee23,rinaldi25a,juodzbalis24_rosetta,labbe24} Thus likely hinting at a presence of the host-galaxy in the filters covering rest-UV. In each panel we show the total AB magnitude as presented in the GLIMPSE catalog \citep{kokorev25a,atek25}. The source is exceptionally bright (M444$\sim23.6$) and is detected in most \jwst\ bands at $>100\sigma$.
\textbf{Middle:} 2D MSA G395M spectra covering \srcname.
\textbf{Bottom:} Combined 1D spectrum \citep[for extraction method see e.g.][]{kokorev23c,de_graaff24} of the LRD in the observed frame. We show the data in black, and the uncertainty with a black shaded region.
Fixing the systemic redshift to the $[$S\,\textsc{iii}$]$ $\lambda$9071 line -  $z_{\rm spec}$ = $3.50102\pm0.00019$ - we show the positions and label the prominent emission with significant ($\geq3\sigma$) detections as solid vertical lines. Iron lines are shown separately in green. Due to the sheer number of features, not all could be labeled, we show all of them in \autoref{fig:fig_lines}. Emission lines for which we only obtain an upper limit are shown with dashed lines.}
\label{fig:fig1}
\end{figure*}

\section{Observations and Data} \label{sec:obs_data}
The target - \srcname\ was originally identified as a bright LRD candidate in the \jwst\ NIRCam imaging captured by the GLIMPSE (PID: 3293; PIs: H. Atek \& J. Chisholm) survey \citep{atek25} of the lensed AS1063 Hubble Frontier Field \citep{lotz17}. The highly magnified area of AS1063 has already successfully yielded $z>16$ galaxy candidates \citep{kokorev25a}, potential Population III hosts \citep{fujimoto25_pop3}, numerous faint and high-redshift galaxies \citep{Chemerynska}, new constraints on reionization \citep{korber}, identification of intermediate mass black-holes \citep{fei25}, and a wide variety of enigmatic LRDs, some moderately lensed. The LRD selection was based on the standard compactness plus ``v-shape" criteria already laid out in \citet{labbe23,greene24,akins24,kokorev24}, using the photometric redshifts derived with \textsc{eazy} \citep{brammer08}, a technique that has proven successful at consistently identifying many exciting sources \citep[see e.g.][]{taylor25,kokorev23c,naidu25_bh*,degraaff25_cliff,akins25}. The LRD in question - \srcname\, located at $z_{\rm phot}=4.2\pm0.1$, stood out in particular due to its extreme rest-optical brightness with $m_{\rm f444w}\sim23.6$ mag and a modest lensing magnification $\mu\sim2$. Very little could be done with photometry alone, however this marked \srcname, as a promising target for any future spectroscopic follow-up. 

\subsection{Photometry}
\label{sec:phot}
We use photometry both to select targets for the GLIMPSE-D NIRSpec observations and, during spectroscopic modeling, to constrain the overall shape of the SED and correct for NIRSpec slit losses. In addition to \jwst/NIRCam data, we incorporate deep \textit{HST} ACS and WFC3 imaging from the Hubble Frontier Fields \citep[HFF;][]{lotz17} and BUFFALO \citep{steinhardt20} programs. Our reprocessed \textit{HST} mosaics are based on Gaia-aligned images from the CHArGE archive \citep{kokorev22}, hosted on the Dawn JWST Archive \citep[DJA;][]{valentino23}. More descriptions of the image reduction and source extraction procedure can be found in \citet{endlsey24}, as well as in the GLIMPSE overview paper \citep{atek25}, we briefly summarize the latter procedure below.

Photometry is performed on PSF-homogenized \textit{HST} and \jwst\ images, convolved to the resolution of the F480M filter. Source detection is carried out using \textsc{SExtractor} \citep{sextractor} through two parallel steps. We construct two inverse-variance-weighted detection images: one from the short-wavelength (SW; F090W, F115W, F150W, F200W) bands to preserve spatial resolution, and one from the long-wavelength (LW; F277W, F356W, F444W) bands to ensure sensitivity to red or dusty sources (e.g., LRDs). Unlike the science images, these have not been PSF matched. These SW and LW detection catalogs are subsequently merged into a single, combined catalog. Photometry is then measured using \textsc{photutils} \citep{photutils} in a range of circular apertures ($D = 0\farcs1$–$1\farcs2$). Photometric uncertainties are estimated by placing random apertures in empty regions around each source. Unless stated otherwise, we adopt total fluxes measured within a $D=0\farcs2$ aperture throughout this work.

\subsection{NIRSpec Observations}
\label{sec:data}
GLIMPSE DDT (GLIMPSE-D hereafter) NIRSpec data were obtained during a campaign (DDT \#9223 PIs: S. Fujimoto \& R. Naidu) to study a promising Population III galaxy candidate \citep{fujimoto25_pop3}. GLIMPSE-D obtained a total of 3 
G395M/F295LP MSA pointings, totaling 29.78h of total exposure time. The pointing center was different for all three configurations to maximize the total object yield. In total,the GLIMPSE-D sample contains 384 spectra with depths varying from 9.2h to $\sim30$h. All of the planned observations were successfully executed between June 30 and July 2 2025. For each MSA configuration, GLIMPSE-D employed a standard 3-point nod pattern at an aperture position angle \texttt{PA\_V3} = 270.5$^{\circ}$, using 494 groups per integration with the NRSIRS2 readout mode. The full details of the target selection, prioritization and MSA planning will be presented in a forthcoming survey paper (Fujimoto. S \& Naidu. R., et al in prep.).

\subsection{G395M Data Reduction and Calibration}
\label{sec:data_red}
The GLIMPSE-D MSA spectra were uniformly reduced using \textsc{msaexp} \citep[v0.9.8;][]{msaexp}. This procedure starts with the level-2 calibrated products obtained from MAST and applies a number of corrections that include 1/f noise, artifact detection and removal and a bias correction in individual exposures \citep[see e.g.][for more details]{rigby23}. Together with the \jwst\ pipeline \textsc{msaexp} sets the slit WCS, performs flat-fielding and computes an initial pass-loss correction. Each one of the 2D shutters in then drizzled onto a common pixel grid. Using the standard approach, the background subtraction is performed locally by using the stacked, source-free shutters. The 1D spectra are then obtained by using the optimal extraction method \citep[e.g.][]{de_graaff23,arrabal_haro23}, where the center and width of the extraction ``aperture'' varies depending on the best fit Gaussian model \citep{horne86}; similar to the methodology which has been widely adopted by a variety of other works \citep[e.g.][]{wang23,greene24,kokorev24b}.

The absolute flux calibration of MSA spectra can be influenced by several factors, including the position of the source within the shutter, calibration and astrometric uncertainties, and the intrinsic morphology of the source. To correct for these effects and derive an overall slit-loss correction, we rescale the extracted 1D spectra by convolving them with all available NIRCam filters and comparing the resulting flux densities to the total photometry from the GLIMPSE catalog. A wavelength-dependent correction is then obtained by fitting a second-order polynomial to these differences.

The target in this paper was observed in all three configurations, however one of them is severely contaminated by sources in the same row, making the data recovery for this source unfeasible. Despite that, \srcname\ is securely detected in two configurations (33262s and 40703s), making the total integration time 20.55h. We combine the extracted and separately calibrated 1D spectra by using inverse-variance weighting. The stacked spectrum is shown in \autoref{fig:fig1}. We also make the spectrum publicly available \footnote{\url{https://zenodo.org/records/17575391}}

\begin{figure*}[h]
\begin{center}
\includegraphics[width=.95\textwidth]{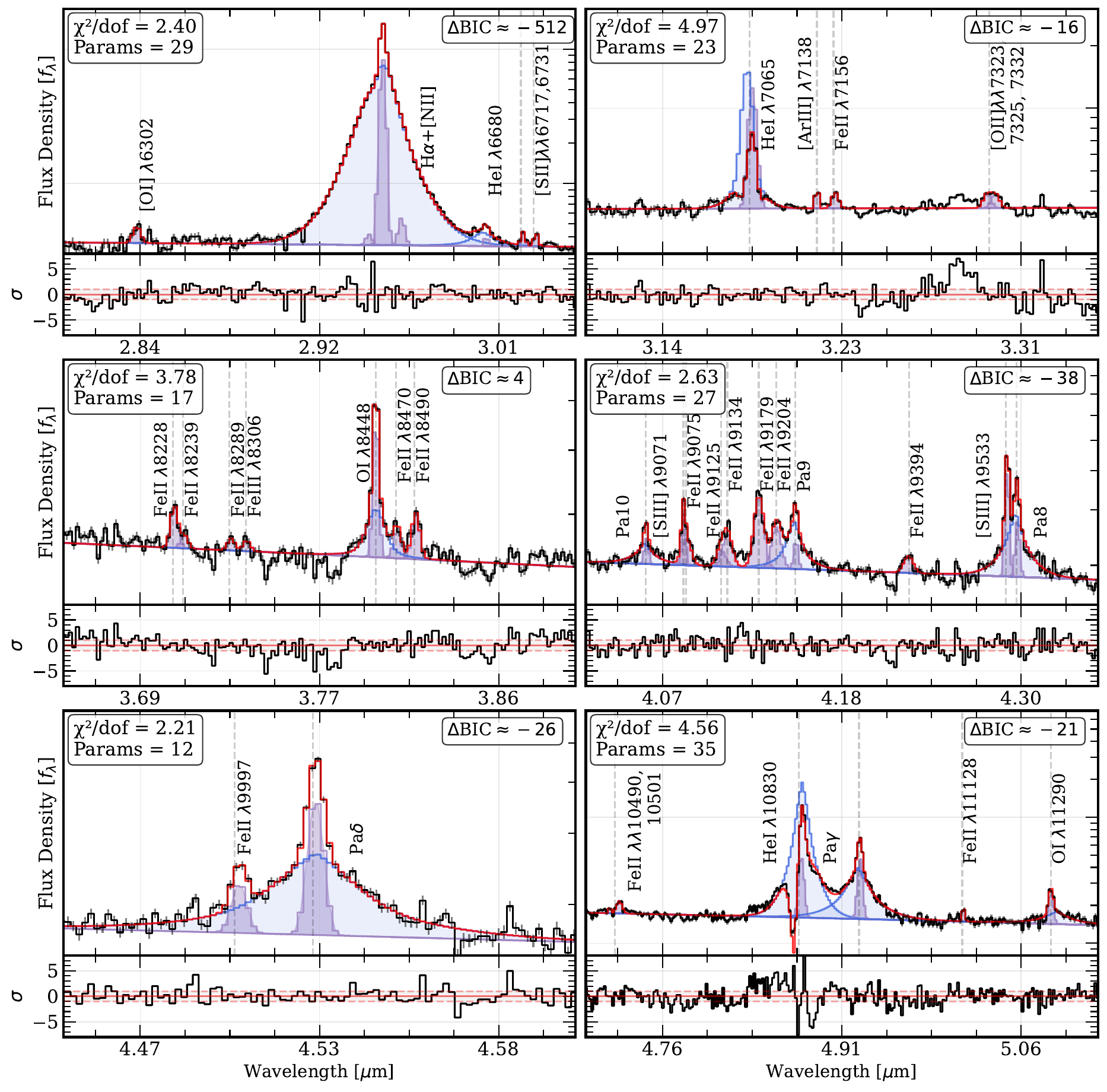}
\caption{\textbf{A staggering abundance of spectral lines in \srcname\, at z=3.501.}
For each spectral window defined in \autoref{sec:lines}, we show the data (black), the best-fit narrow and broad components (dark purple and blue), and the total best-fit model including the continuum (red). All broad components were fit with models allowing exponential wings. The $\Delta$BIC between exponential and Gaussian fits is reported in the top right of each panel, negative number indicates a preference for an exponential profile. The lower panels display the uncertainty-weighted residuals for each fit.}
\label{fig:fig_lines}
\end{center}
\end{figure*}

\begin{figure*}[t]
\begin{center}
\includegraphics[width=.95\textwidth]{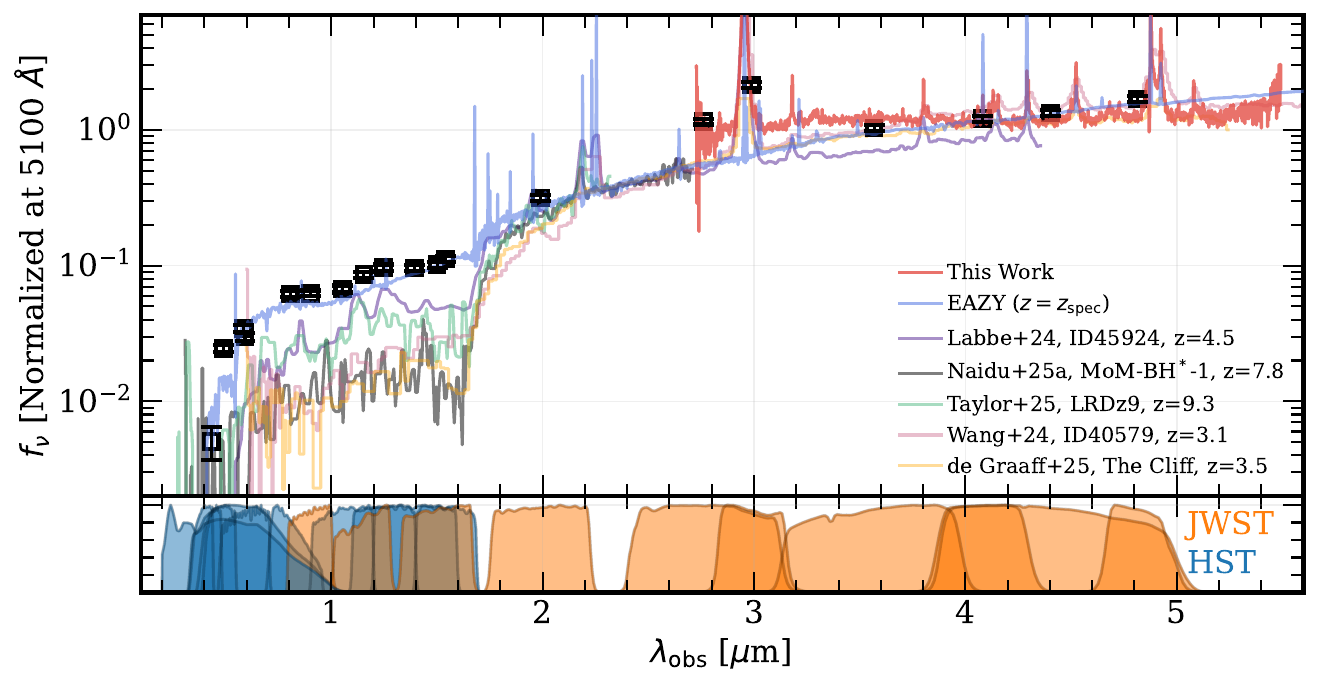}
\caption{\textbf{The diversity of Balmer breaks in LRDs.} Black points show the \textit{HST} and \jwst\ GLIMPSE photometry of \srcname. Blue line shows best-fit \textsc{EAZY} SED fit to the photometry only, fixing the redshift to the $z_{\rm spec}$. Maroon line show the combined and photometry-corrected G395M spectrum. While the red color in F200W-F277W is partially influenced by a bright H$\alpha$ line, the Balmer break between F150W and F200W is still prominent. We further show spectra of various other LRDs \citep{labbe24,naidu25_bh*,taylor25,wang24,degraaff25_cliff}, all shifted to $z=3.501$ and normalized at 5100 \AA. Finally, we show \textit{HST} (blue) and \jwst\ (orange) filter transmission curves below.} 
\label{fig:fig_sed}
\end{center}
\end{figure*}

\section{Emission Line Analysis}
\label{sec:emline_analysis}

\subsection{Spectroscopic Redshift}
\label{sec:zspec}
The spectrum of \srcname\ reveals a staggering wide variety of significantly detected emission and absorption lines. Before focusing on individual features and their complex components, we perform an initial tally of all detectable lines using a heavily modified version of \textsc{msaexp} \citep{msaexp,kokorev24b}. The key modifications allow the ability to vary the line widths, as well as fit multiple components, albeit tied to the same redshift.

We fit the full spectrum using \textsc{msaexp}, adopting Gaussian profiles for emission lines and a three-segment cubic spline to model the continuum. This minimal spline structure, appropriate given the high average S/N of \srcname\ ($\gtrsim$30), helps prevent overfitting, particularly of features that may be slightly offset in wavelength/velocity space from the average redshift. Emission line positions are fixed; narrow components are allowed FWHM between 150–800 km s$^{-1}$, and permitted transitions may include a broad component (800–5000 km s$^{-1}$).

This yields a redshift of $z_{\rm spec} = 3.5010 \pm 0.0001$. However, several strong lines exhibit significant pixel-level offsets of 100–200 km s$^{-1}$ relative to this value. These discrepancies are statistically robust and appear in both stacked and individual spectra, suggesting that the derived redshift reflects a weighted average dominated by high-S/N lines. To properly assess velocity structure, a consistent systemic redshift is required. This will be explored in the following sections. 

Using the updated redshift alongside the strong-lensing model of AS1063 \citep{zitrin15} and L. Furtak et al. (in prep.), we recalculate the lensing magnification to be $\mu = 2.04 \pm 0.21$, consistent with previous estimates based on $z_{\rm phot}$. With redshift and magnification now fixed, we turn to a deeper analysis of the emission line properties.

\subsection{Line Identification}
\label{sec:line_ident}
Beyond velocity offsets, the initial \textsc{msaexp} fit highlights several notable features. Broad hydrogen lines, which are ubiquitous in LRDs \citep[e.g.][]{hviding25,matthee23,kocevski23,kokorev23c,furtak24,taylor25,degraaff25_cliff,naidu25_bh*}, are strongly favored across the Balmer and Paschen series, with $\Delta\chi^2 > 10$. These include H$\alpha$ and Pa10 through Pa$\gamma$. Although Pa$\beta$ falls near the edge of the detector, we nonetheless observe a prominent broad wing at its expected location (see \autoref{fig:fig1}).

Broad components are also detected in several helium and oxygen lines: He\,\textsc{i}$\,{\lambda6680}$, He\,\textsc{i}$\,{\lambda7065}$, O\,\textsc{i}$\,{\lambda8448}$, He\,\textsc{i}$\,{\lambda10830}$, and O\,\textsc{i}$\,{\lambda11290}$. Notably, the He\,\textsc{i}$\,{\lambda10830}$ line shows a classic P-Cygni profile with a clear blueshifted absorption trough. While Balmer absorption is becoming a familiar sight in LRDs \citep{matthee23, maiolino23b, kocevski24, degraaff25_cliff, naidu25_bh*}, helium absorption remains rare, with only a few published cases \citep[e.g.][]{naidu24,wang25, juodzbalis24_rosetta} at high-$z$. Curiously this was also recently reported in local analogs of LRDs \citep{sdss_lrds}. 

The most striking feature is the detection of an extensive Fe\,\textsc{ii} emission forest in the rest-NIR—at least 16 distinct features are identified by \textsc{msaexp} and highlighted in green in \autoref{fig:fig1}. Previous detections of permitted Fe\,\textsc{ii} in LRDs have been mostly limited to rest-UV/optical wavelengths using low-resolution PRISM spectra \citep{tripodi25}, with only few examples in the NIR \citep[e.g. see broad Fe\,\textsc{ii} $\lambda9200$][]{labbe24}. Iron has also been identified in local LRD analogs \citep{sdss_lrds,ji25}. Only recently have a few Fe\,\textsc{ii} emitters been reported in medium- \citep{deugenio25} and high-resolution spectra \citep{torralba25}. The richness of these lines in \srcname\ hints at a dense, partially shielded gas phase, where processes such as Ly$\beta$ fluorescence and continuum pumping may be enhancing Fe\,\textsc{ii} emission \citep{sigut98,sigut03,sigut04}, offering a new diagnostic window into early AGN environments.

In summary, \srcname\ exhibits an exceptionally rich and kinematically complex array of emission and absorption features. While the initial \textsc{msaexp} modeling provides a solid foundation for redshift estimation and line identification, it lacks the flexibility to capture the full diversity of line profiles in this high-S/N dataset. In the following sections, we peel back the spectral layers with increasing precision.


\subsection{Line Fitting}
\label{sec:lines}
The exceptional depth of the G395M spectrum for \srcname\ enables detailed modeling of both multiple kinematic components and potential velocity offsets between line species. Unless noted otherwise, each line is fit with a single Gaussian narrow component (FWHM = 100–800 km s$^{-1}$). For permitted lines flagged by \textsc{msaexp} as potentially broad, we model the core with a Gaussian convolved with an exponential tail \citep{rusakov25}, with FWHM allowed between 500–5000 km s$^{-1}$. The degree of the convolution, i.e. the exponential ``strength'' is allowed to vary freely, however for every broad line we also perform a fit with that parameter fixed to zero (i.e. pure Gaussian) and compare it to our free fit. Line centers are allowed to vary independently, unless otherwise specified.

Absorption lines are modeled using a standard attenuation law with optical depth ($\tau_\nu$) as a free parameter \citep[see][]{juodzbalis24_rosetta}. Widths and redshifts (for both narrow and broad components) are typically free, with redshift limited to $\pm0.1$ from the \textsc{msaexp} value. Local continua are fit using first-order polynomials.

All of the individual model components are first initialized and co-added on an oversampled wavelength grid. To take into account the wavelength dependent resolution of the grating, we interpolate our model onto a variable step grid while making sure that the total integrated flux is preserved. Further,  we increase the nominal spectral resolution by a factor of 1.7, as it has been shown that the spectral resolution for a point-like source falling within a shutter is higher than that of a uniformly illuminated slitlet \citep{de_graaff23}. Fitting uses nonlinear $\chi^2$ minimization with uncertainties being derived from multivariate resampling of the covariance matrix.

Given the complexity, the spectrum is divided into six windows grouped by line species or proximity (see \autoref{fig:fig_lines}). Below we describe assumptions per window. Final fluxes and equivalent widths of the narrow and broad lines are listed in \autoref{tab:emission_lines} and \autoref{tab:emission_lines_broad}, respectively, while kinematics are reported in \autoref{tab:kinematics_narrow} and \autoref{tab:kinematics_broad}. We show detailed line fits in \autoref{fig:fig_lines}.

\subsubsection{H\texorpdfstring{$\alpha$}{alpha} complex}
H$\alpha$ is modeled with independent narrow, broad, and absorption components. [O\,\textsc{i}]$\,{\lambda6302}$ and He\,\textsc{i}$\,{\lambda6680}$ are modeled with narrow and broad profiles. [N\,\textsc{ii}] lines are fixed at a 1:3 ratio and share kinematics; [S\,\textsc{ii}] lines have independent amplitudes but tied velocities and widths.

We find that broad H$\alpha$ and He\,\textsc{i}$\,{\lambda6680}$ are strongly favored to have exponential wings ($\Delta$BIC $\sim -500$), yielding a FWHM (of the intrinsic Gaussian core) of $\sim 1000$ km s$^{-1}$—significantly narrower than the $\sim 3000$ km s$^{-1}$ derived from pure Gaussians. Although no distinct H$\alpha$ absorption component is explicitly detected, likely due to insufficient resolution, the overall asymmetry of the line is best reproduced when a weak, blueshifted absorber is included in the fit. This may indicate subtle self-absorption or partial obscuration within the dense cocoon. We return to the implications of this profile shape later.

\subsubsection{He\,\textsc{i}\,${\lambda7065}$}
The He\,\textsc{i}$\,{\lambda7065}$ line shows a prominent blueshifted absorption component. We fit it with narrow, broad, and absorption profiles. The rest of the lines which include [Ar\,\textsc{iii}]$\,{\lambda7138}$, Fe\,\textsc{ii}$\,{\lambda7156}$, and the [O\,\textsc{ii}] triplet ($\lambda\lambda7323,7325,7332$) are modeled as independent narrow lines.

He\,\textsc{i}$\,{\lambda7065}$ shows a strong preference for an exponential profile ($\Delta$BIC $\sim -16$) with FWHM = $473 \pm 116$ km s$^{-1}$. A residual bump spanning from $\sim\lambda7290$--7320 is likely a combination of multiple 
Fe\,\textsc{ii}, such as 7290, 7308 and potentially Fe\,\textsc{iii}$\,{\lambda7319.65}$ lines. Curiously the latter line is often labeled as ``hazy'' in various emission line libraries \citep[e.g.][]{nist}, which has an undefined shape as a result of pressure broadening or scattering in a very dense gas.

\subsubsection{O\,\textsc{i}$\,{\lambda8448}$ and Iron Lines}
In this part of the spectrum the O\,\textsc{i}$\,{\lambda8448}$ is fit with narrow and broad components. All of the iron lines, which include Fe\,\textsc{ii}$\,{\lambda \lambda8228,8239,8289,8470,8490}$
as well as a potential Fe\,\textsc{iii}$\,{\lambda8306}$, are fit as narrow lines with shared kinematics. O\,\textsc{i} components are fit independently.

We find that the broad O\,\textsc{i} component is best described by a standard Gaussian profile, with a stronger statistical preference over exponential wings ($\Delta$BIC $\sim 4$). However, we caution that a trough at $\sim$3.77\,$\mu$m, likely an artifact, may affect the reliability of the fit in this region

\subsubsection{Iron Forest and Paschen Lines}
This spectral region is among the most complex, due to the dense clustering of emission features. The Paschen lines Pa10–Pa8 are modeled with narrow and broad components. To reduce the number of free parameters, we tie all narrow Paschen components together kinematically, and do the same for the broad components. Fe\,\textsc{ii}$\,{\lambda\lambda9075,9125,9134,9179,9204,9394}$ are fit with shared centroid and width. [S\,\textsc{iii}] lines are fit independently.

The [S\,\textsc{iii}]$\,{\lambda9071}$ is used to define the systemic redshift: $z = 3.50102 \pm 0.00019$. This is done because the forbidden lines, such as [S\,\textsc{iii}] or [O\,\textsc{iii}], have lower critical densities, incompatible with the dense gas envelopes that gives rise to other line we observe. These lines may instead come from the host galaxy itself \citep[e.g.][]{maiolino23b} and thus we choose this line to define the overall redshift of the system to measure all the offsets relative to it. All velocity offsets discussed in the subsequent sections are computed relative to this reference frame. The Paschen lines show strong preference for exponential wings ($\Delta$BIC $\sim -38$), with FWHM = $766 \pm 202$ km s$^{-1}$—narrower than H$\alpha$, but consistent within $2\sigma$.

\subsubsection{Pa$\delta$}
The Fe\,\textsc{ii}$\,{\lambda9997}$ and Pa$\delta$ complex is modeled with an independent narrow component for Fe\,\textsc{ii} and both narrow and broad components for Pa$\delta$. There is no evidence for absorption. Exponential wings are again preferred ($\Delta$BIC $\sim -26$). The resulting FWHM of $1672 \pm 458$ km s$^{-1}$ is notably larger than that of both higher-order Paschen lines and H$\alpha$, though still consistent with the latter within 2$\sigma$. It is also possible that the shape is impacted by the Fe\,\textsc{ii}$\,{\lambda9997}$ and the undetected (but likely present) Fe\,\textsc{ii}$\,{\lambda\lambda 10131, 10173}$ lines.

\subsubsection{He\,\textsc{i}$\,{\lambda10830}$ and Pa$\gamma$}
The final window contains a complicated blend of broad He\,\textsc{i}$\,{\lambda10830}$ and Pa$\gamma$ lines, with a clear blueshifted absorption in the former. We fit a three component (narrow, broad and absorption) to He\,\textsc{i}${\lambda10830}$ and narrow+broad component to Pa$\gamma$, keeping everything kinematically independent. We fit narrow Fe\,\textsc{ii} at 10490, 10501 and 11128 \AA\, with fixed redshift and line width. Finally the O\,\textsc{i}$\,{\lambda11290}$ is fit with independent narrow and broad component. Exponential wings are again preferred over pure Gaussian profile.

\begin{figure}
\begin{center}
\includegraphics[width=.49\textwidth]{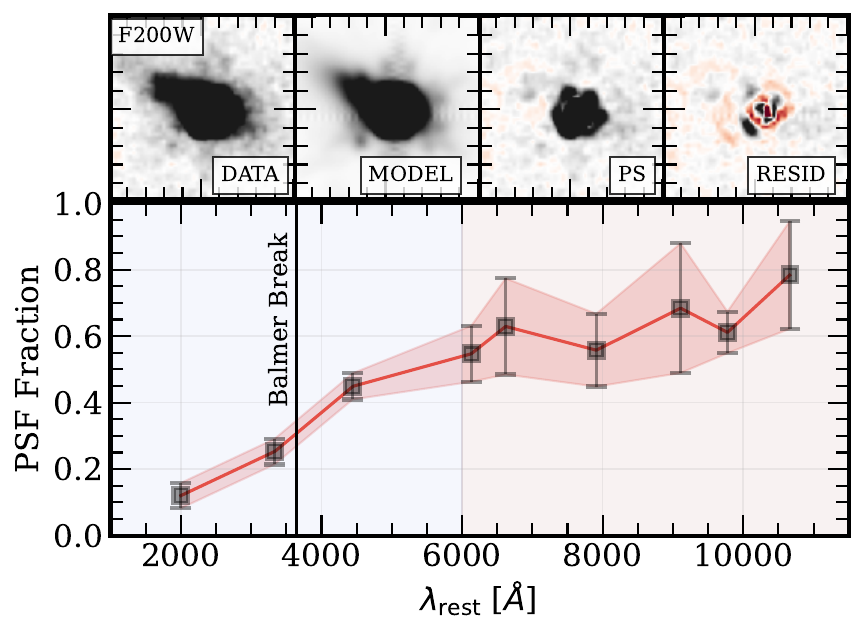}
\caption{\textbf{Top:} Two-component S\'ersic + point-source fit to the F200W morphology of \srcname. Panels show, from left to right, the data, best-fit model, model with the extended component removed, and residuals. \textbf{Bottom:} Fractional point-source contribution versus rest-frame wavelength. Shaded regions mark SW (blue) and LW (red) detectors, where the lower resolution of LW filters hinders reliable two-component decomposition.} 
\label{fig:fig_morph}
\end{center}
\end{figure}

\begin{deluxetable}{cc}[]
\tabcolsep=2mm
\tablecaption{\label{tab:tab1} Source Properties. }.
\tablehead{Parameter$^\dagger$ &  \srcname}
\startdata
ID & 17775 \\
RA [deg] & 342.20080 \\
Dec [deg] & -44.54366\\
$z_{\rm phot}$ (\textsc{eazy}) &  $4.2\pm0.1$ \\
$z_{\rm spec}$ ($[$S\,\textsc{iii}$]$ $\lambda9071$) & $3.50102\pm0.00019$  \\
$\mu$ & $2.04\pm0.21$\\
$M_{\rm UV}$ [AB mag] & $-17.27\pm0.05$\\
$r_{\rm eff,UV}$ [pc] & 1000$\pm$200\\
$r_{\rm eff,opt}$ [pc] & $<300$\\
$\beta$ & -0.69$\pm$0.12 \\
log$_{10}$($M_{\rm BH}/M_\odot$) & 6.65$\pm$0.15  \\
$L_{\rm bol}$ [erg/s] & (1.06$\pm$0.14)$\times10^{45}$  \\
$\lambda_{\rm edd}$ & $1.86\pm0.25$  \\
log$_{10}$($M_*/M_\odot$)$_{\rm M_{\rm UV}}$ & $<7.5$  \\
$A_{\rm V}$ & 0.1$\pm$0.3  \\
$f_{\nu,4050\textrm{\AA{}}}/f_{\nu,3670\textrm{\AA{}}}$ & $2.02\pm0.10$ \cr
\enddata
\begin{tablenotes}
\footnotesize{$^\dagger$ Physical parameters are corrected for the lensing magnification.} \\
\end{tablenotes}
\end{deluxetable}

\begin{deluxetable}{cccc}[]
\tabcolsep=2mm
\tablecaption{\label{tab:emission_lines}
Fluxes$^\dagger$ of narrow emission lines and their rest-frame equivalent widths.}
\tablehead{
\colhead{Line} & 
\colhead{$\lambda_{\rm rest}$} & 
\colhead{Flux} & 
\colhead{EW$_0$} \\
\colhead{} & 
\colhead{[\AA]} & 
\colhead{[10$^{-20}$ erg s$^{-1}$ cm$^{-2}$]} & 
\colhead{[\AA]}
}
\startdata
\multicolumn{4}{c}{\textbf{Narrow Emission Lines}} \\
\hline
O\,\textsc{i} & 6302.0 & $54.0 \pm 21.2$ & $3.3 \pm 1.2$ \\
$[$N\,\textsc{ii}$]$ & 6549.0 & $25.6 \pm 10.3$ & $1.6 \pm 0.7$ \\
H$\alpha$ & 6562.8 & $2622.9 \pm 200.7$ & $167.7 \pm 12.8$ \\
$[$N\,\textsc{ii}$]$ & 6584.0 & $77.1 \pm 37.1$ & $4.9 \pm 1.6$ \\
He\,\textsc{i} & 6680.0 & $14.7 \pm 11.1$ & $0.8 \pm 0.7$ \\
$[$S\,\textsc{ii}$]$ & 6717.0 & $22.2 \pm 0.1$ & $1.5 \pm 0.7$ \\
$[$S\,\textsc{ii}$]$ & 6731.0 & $21.4 \pm 0.1$ & $1.4 \pm 0.5$ \\
He\,\textsc{i} & 7065.0 & $461.7 \pm 210.2$ & $31.0 \pm 12.2$ \\
$[$Ar\,\textsc{iii}$]$ & 7138.0 & $12.6 \pm 3.4$ & $1.0 \pm 0.3$ \\
Fe\,\textsc{ii} & 7156.0 & $26.2 \pm 7.2$ & $1.8 \pm 0.2$ \\
$[$O\,\textsc{ii}$]$ & 7323.0 & $26.0 \pm 23.0$ & $1.7 \pm 1.5$ \\
$[$O\,\textsc{ii}$]$ & 7325.0 & $16.9 \pm 1.8$ & $1.2 \pm 0.1$ \\
$[$O\,\textsc{ii}$]$ & 7332.0 & $26.2 \pm 9.1$ & $1.8 \pm 0.6$ \\
Fe\,\textsc{ii} & 8228.0 & $23.7 \pm 4.2$ & $2.0 \pm 0.3$ \\
Fe\,\textsc{ii} & 8239.0 & $7.2 \pm 4.0$ & $0.6 \pm 0.3$ \\
Fe\,\textsc{ii} & 8289.0 & $6.3 \pm 3.7$ & $0.5 \pm 0.3$ \\
Fe\,\textsc{iii} & 8306.0 & $5.7 \pm 3.6$ & $0.5 \pm 0.3$ \\
O\,\textsc{i} & 8448.0 & $67.6 \pm 2.4$ & $6.1 \pm 1.3$ \\
Fe\,\textsc{ii} & 8470.0 & $16.2 \pm 5.0$ & $1.4 \pm 0.4$ \\
Fe\,\textsc{ii} & 8490.0 & $26.7 \pm 4.3$ & $2.4 \pm 0.4$ \\
Pa10 & 9015.0 & $16.5 \pm 5.0$ & $1.8 \pm 0.5$ \\
$[$S\,\textsc{iii}$]$ & 9071.0 & $27.2 \pm 2.3$ & $3.5 \pm 0.3$ \\
Fe\,\textsc{ii} & 9075.0 & $30.0 \pm 4.7$ & $3.4 \pm 0.5$ \\
Fe\,\textsc{ii} & 9125.0 & $16.8 \pm 5.2$ & $1.9 \pm 0.6$ \\
Fe\,\textsc{ii} & 9134.0 & $36.9 \pm 5.0$ & $4.1 \pm 0.6$ \\
Fe\,\textsc{ii} & 9179.0 & $95.0 \pm 5.5$ & $10.6 \pm 0.6$ \\
Fe\,\textsc{ii} & 9204.0 & $50.1 \pm 6.1$ & $5.6 \pm 0.7$ \\
Pa9 & 9229.0 & $19.4 \pm 7.0$ & $2.2 \pm 0.8$ \\
Fe\,\textsc{ii} & 9394.0 & $17.3 \pm 3.9$ & $2.0 \pm 0.5$ \\
$[$S\,\textsc{iii}$]$ & 9533.0 & $84.7 \pm 5.9$ & $10.2 \pm 0.7$ \\
Pa8 & 9545.0 & $44.8 \pm 11.3$ & $5.4 \pm 1.4$ \\
Fe\,\textsc{ii} & 9997.0 & $29.5 \pm 4.5$ & $3.8 \pm 0.6$ \\
Pa$\delta$ & 10049.0 & $104.2 \pm 12.2$ & $13.7 \pm 1.7$ \\
Fe\,\textsc{ii} & 10490.0 & $8.0 \pm 2.7$ & $1.1 \pm 0.4$ \\
Fe\,\textsc{ii} & 10501.0 & $24.4 \pm 5.9$ & $3.1 \pm 0.8$ \\
He\,\textsc{i} & 10830.0 & $163.0 \pm 96.0$ & $22.6 \pm 13.3$ \\
Pa$\gamma$ & 10938.0 & $146.3 \pm 25.5$ & $20.9 \pm 3.8$ \\
Fe\,\textsc{ii} & 11128.0 & $12.3 \pm 3.7$ & $1.8 \pm 0.6$ \\
O\,\textsc{i} & 11290.0 & $49.2 \pm 4.1$ & $7.7 \pm 2.3$ \\
\enddata
\begin{tablenotes}
\footnotesize{$^\dagger$ Not corrected for lensing magnification.} \\
\end{tablenotes}
\end{deluxetable}

\begin{deluxetable}{cccc}[]
\tabcolsep=2mm
\tablecaption{\label{tab:emission_lines_broad}
Fluxes$^\dagger$ of broad emission lines and their rest-frame equivalent widths.}
\tablehead{
\colhead{Line} & 
\colhead{$\lambda_{\rm rest}$} & 
\colhead{Flux} & 
\colhead{EW$_0$} \\
\colhead{} & 
\colhead{[\AA]} & 
\colhead{[10$^{-20}$ erg s$^{-1}$ cm$^{-2}$]} & 
\colhead{[\AA]}
}
\startdata
\multicolumn{4}{c}{\textbf{Broad Emission Lines}} \\
\hline
H$\alpha$ & 6562.8 & $14803.5 \pm 198.6$ & $946.3 \pm 15.6$ \\
He\,\textsc{i} & 6680.0 & $144.6 \pm 44.7$ & $9.4 \pm 2.9$ \\
He\,\textsc{i} & 7065.0 & $824.2 \pm 210.0$ & $55.4 \pm 15.1$ \\
Pa10 & 9015.0 & $53.7 \pm 11.1$ & $5.8 \pm 1.3$ \\
Pa9 & 9229.0 & $119.0 \pm 13.5$ & $13.6 \pm 1.5$ \\
Pa8 & 9545.0 & $191.2 \pm 17.0$ & $23.1 \pm 2.1$ \\
Pa$\delta$ & 10049.0 & $370.6 \pm 15.0$ & $48.8 \pm 2.3$ \\
He\,\textsc{i} & 10830.0 & $2294.9 \pm 212.6$ & $318.8 \pm 30.3$ \\
Pa$\gamma$ & 10938.0 & $652.3 \pm 14.0$ & $92.9 \pm 2.4$ \\
O\,\textsc{i} & 8448.0 & $93.7 \pm 7.9$ & $8.4 \pm 0.7$ \\
O\,\textsc{i} & 11290.0 & $106.2 \pm 63.9$ & $16.6 \pm 10.4$ \\
\enddata
\begin{tablenotes}
\footnotesize{$^\dagger$ Not corrected for lensing magnification.} \\
\end{tablenotes}
\end{deluxetable}

\begin{deluxetable}{cccc}
\label{tab:kinematics_narrow}
\tabcolsep=2mm
\tablecaption{Kinematic properties of narrow lines.}
\tablehead{
\colhead{Line} & \colhead{$\lambda_{\rm rest}$} & \colhead{FWHM} & \colhead{$\Delta v$} \\
\colhead{} & \colhead{[\AA]} & \colhead{[km s$^{-1}$]} & \colhead{[km s$^{-1}$]}
}
\startdata
\multicolumn{4}{c}{\textbf{Narrow Emission Lines}} \\
\hline
$[$O\,\textsc{i}$]$ & 6300.3 & $455\pm175$ & $-101 \pm 81$ \\
H$\alpha$ & 6562.8 & $304 \pm 8 $ & $+145 \pm 13$ \\
$[$N\,\textsc{ii}$]$ & 6583.4 & $347 \pm 117$ & $+84 \pm 47$ \\
$[$S\,\textsc{ii}$]$ & 6716.4 & $232 \pm 62$ & $+82 \pm 33$ \\
$[$S\,\textsc{ii}$]$ & 6731 & $232 \pm 62$ & $+82 \pm 33$ \\
He\,\textsc{i} & 6678.2 & $87 \pm 14$ & $+40 \pm 14$ \\
He\,\textsc{i} & 7065.2 & $408 \pm 644$ & $+112 \pm 396$ \\
$[$Ar\,\textsc{iii}$]$ & 7135.8 & $89 \pm 50$ & $-26 \pm 20$ \\
Fe\,\textsc{ii} & 7155.2 & $96 \pm 50$ & $+15 \pm 15$ \\
$[$O\,\textsc{ii}$]$ & 7320.0 & $400 \pm 120$ & $-37 \pm 72$ \\
Fe\,\textsc{ii} & 8228.0 & $290 \pm 48$ & $+21 \pm 5$ \\
Fe\,\textsc{ii} & 8239.0 & $290 \pm 48$ & $+21 \pm 5$ \\
Fe\,\textsc{ii} & 8289.0 & $290 \pm 48$ & $+21 \pm 5$ \\
Fe\,\textsc{iii} & 8306.0 & $290 \pm 48$ & $+21 \pm 5$ \\
O\,\textsc{i} & 8446.4 & $235 \pm 30$ & $-43 \pm 19$ \\
Fe\,\textsc{ii} & 8470.0 & $290 \pm 48$ & $+21 \pm 5$ \\
Fe\,\textsc{ii} & 8490.0 & $290 \pm 48$ & $+21 \pm 5$ \\
Pa10 & 9014.9 & $259 \pm 36$ & $-46 \pm 19$ \\
$[$S\,\textsc{iii}$]$ & 9071.0 & $230 \pm 13$ & -- \\
Fe\,\textsc{ii} & 9075.0 & $486 \pm 24$ & $-21 \pm 17$ \\
Fe\,\textsc{ii} & 9125.0 & $486 \pm 24$ & $-21 \pm 17$ \\
Fe\,\textsc{ii} & 9134.0 & $486 \pm 24$ & $-21 \pm 17$ \\
Fe\,\textsc{ii} & 9179.0 & $486 \pm 24$ & $-21 \pm 17$ \\
Fe\,\textsc{ii} & 9204.0 & $486 \pm 24$ & $-21 \pm 17$ \\
Pa9 & 9229.0 & $259 \pm 35$ & $-46 \pm 19$ \\
Fe\,\textsc{ii} & 9394.0 & $486 \pm 24$ & $-21 \pm 17$ \\
Pa8 & 9545.6 & $259 \pm 36$ & $-46 \pm 19$ \\
$[$S\,\textsc{iii}$]$ & 9533 & $230 \pm 13$ & $+23 \pm 13$ \\
Fe\,\textsc{ii} & 9997.0 & $331 \pm 55$ & $+90 \pm 25$ \\
Pa$\delta$ & 10049.4 & $332 \pm 22$ & $+21 \pm 15$ \\
He\,\textsc{i} & 10830.3 & $287 \pm 46$ & $+112 \pm 26$ \\
Pa$\gamma$ & 10938.1 & $267 \pm 25$ & $+26 \pm 17$ \\
Fe\,\textsc{ii} & 10490.0 & $332 \pm 96$ & $+44 \pm 43$ \\
Fe\,\textsc{ii} & 10501.0 & $332 \pm 96$ & $+44 \pm 43$ \\
Fe\,\textsc{ii} & 11128.0 & $153 \pm 116$ & $+37 \pm 30$ \\
O\,\textsc{i} & 11290.0 & $260 \pm 58$ & $-33 \pm 24$ \\
\enddata
\begin{tablenotes}
\footnotesize{Velocity offsets ($\Delta v$) calculated relative to $[$S\,\textsc{iii}$]$ $\lambda$9071 at $z = 3.50102$.} \\
\end{tablenotes}
\end{deluxetable}

\begin{deluxetable}{cccc}
\label{tab:kinematics_broad}
\tabcolsep=2mm
\tablecaption{Kinematic properties of broad and absorption lines.}
\tablehead{
\colhead{Line} & \colhead{$\lambda_{\rm rest}$} & \colhead{FWHM} & \colhead{$\Delta v$} \\
\colhead{} & \colhead{[\AA]} & \colhead{[km s$^{-1}$]} & \colhead{[km s$^{-1}$]}
}
\startdata
\multicolumn{4}{c}{\textbf{Broad Emission Lines}} \\
\hline
H$\alpha$ & 6562.8 & $1024 \pm 21$ & $-7 \pm 16$ \\
He\,\textsc{i} & 6678.2 & $1041\pm792$ & $-137 \pm 151$ \\
He\,\textsc{i} & 7065.2 & $473 \pm 116$ & $-136 \pm 82$ \\
O\,\textsc{i} & 8446.4 & $562 \pm 331$ & $+39 \pm 80$ \\
Pa10 & 9014.9 & $766 \pm 202$ & $-155 \pm 39$ \\
Pa9 & 9229.0 & $766 \pm 202$ & $-155 \pm 39$ \\
Pa8 & 9545.6 & $766 \pm 202$ & $-155 \pm 39$ \\
Pa$\delta$ & 10049.4 & $1672 \pm 458$ & $+23 \pm 44$ \\
He\,\textsc{i} & 10830.3 & $536 \pm 45$ & $+154 \pm 22$ \\
Pa$\gamma$ & 10938.1 & $1035 \pm 322$ & $-75 \pm 31$ \\
O\,\textsc{i} & 11287.0 & $1188 \pm 1099$ & $+245 \pm 214$ \\
\hline
\multicolumn{4}{c}{\textbf{Absorption Lines}} \\
\hline
H$\alpha$ & 6562.8 & $1000 \pm 123$ & $-200 \pm 32$ \\
He\,\textsc{i} & 7065.2 & $587 \pm 1024$ & $-52 \pm 83$ \\
He\,\textsc{i} & 10830.3 & $794 \pm 21$ & $+45 \pm 22$ \\
\enddata
\begin{tablenotes}
\footnotesize{Velocity offsets ($\Delta v$) calculated relative to $[$S\,\textsc{iii}$]$ $\lambda$9071 at $z = 3.50102$.} \\
\end{tablenotes}
\end{deluxetable}

\section{Data Analysis}
\label{sec:data_analysis}

\subsection{Morphology}
\label{sec:morph}
LRDs are unresolved in the rest-optical by definition, with measured sizes consistent with the PSF HWHM ($\lesssim0\farcs07$ in F444W), and in some cases even smaller when dithers align favorably \citep[e.g.][]{labbe23}. Gravitational lensing can further push constraints on their intrinsic sizes to $\lesssim100$ pc \citep{furtak24}.

In the rest-UV, however, a more complex picture is emerging. Several studies have now reported faint, extended, asymmetric components adjacent to the compact core \citep{labbe24,kokorev24b,matthee23,rinaldi25a,rinaldi25b}, often suppressed by surface-brightness dimming. As shown in \autoref{fig:fig1}, \srcname\ likewise consists of two components out to F200W (rest $\sim$4000 Å), coincident with the Balmer break. Although we note that the break itself is much weaker than found in objects with very similar spectra \citep[e.g.][]{wang24} see \autoref{fig:fig_sed}. At longer wavelengths, however, a point-source morphology takes over, consistent with a BH-dominated core. To quantify this transition, we model the NIRCam imaging using \textsc{PYSERSIC} \citep{pasha23} with a minimal configuration: a fixed-center point source plus a freely offset S\'ersic profile. Normalizations, $n$, and $r_{\rm eff}$ are all allowed to vary, and uncertainties are drawn from the MCMC posteriors \citep[e.g.][]{kokorev24}.

Our fits show that the host galaxy dominates up to $\lambda_{\rm rest}\sim4000$~\AA, where the host and nucleus contribute roughly equally, beyond which the point source prevails to $\lambda_{\rm rest}\sim1$~$\mu$m.
Because the NIRCam LW detector has $\gtrsim$2$\times$ poorer resolution than the SW, a clean separation of host and nucleus in the rest-optical is impractical \citep[e.g. see][]{whalen25}. Fortunately, F200W lies at the transition where both the resolution and flux ratio are favorable, so we use that band to illustrate our galaxy/LRD decomposition. 
We illustrate this in \autoref{fig:fig_morph}, which also shows the increasing point-source fraction with wavelength, while acknowledging the uncertainties at longer wavelengths.

After applying the lensing correction, we find that the extended component that dominates the UV light has a radius of $r_{\rm eff}\sim1000$ pc, while the actual compact LRD, dominant in rest-optical, is fully consistent with the PSF and $r_{\rm eff}<300$ pc.

\subsection{Dust Attenuation}
\label{sec:dust}
The ratio between observed emission line fluxes is commonly used to estimate dust extinction. For LRDs, this has typically been achieved through the Balmer decrement, provided that multiple Balmer lines are available. In our case, only H$\alpha$ is detected. While the Paschen series, of which five lines are observed, could, in principle, be used to estimate extinction, the likely presence of stratified dense gas in LRDs \citep{naidu25_bh*,degraaff25_cliff} implies that the intrinsic line ratios may deviate significantly from standard Case B recombination expectations \citep{chang25}. Further, at the wavelengths of the Paschen series, the attenuation curves are practically flat \citep[e.g.][]{calzetti00}, small offsets of the Case B ratios will a significant effect on the derived dust attenuation, but in return require extreme precision on measured line fluxes to be meaningful. Our uncertainties on e.g. Pa$\gamma$/Pa$\delta$ are small, roughly 10-15 \%, this still leads to an impractically high uncertainty on the $A_{\rm V}$. To circumvent this limitation, we instead employ the ratio between the permitted O\,\textsc{i}$\,{\lambda8448}$ and O\,\textsc{i}$\,{\lambda11290}$ lines.
These transitions share a common energy level, and under the assumption of Bowen (Ly$\beta$) Fluorescence, all transitions through the $\lambda11290$ transition also cascade through the $\lambda8448$. This makes their intrinsic intensity set solely by the ratio of the inverse of their wavelengths

To ensure we are comparing the same physical gas component, we isolate the narrow emission peaks in both transitions, which show consistent widths of FWHM $\sim$ 250 km s$^{-1}$. This similarity supports the use of their flux ratio as a dust probe. We adopt the SMC attenuation curve \citep{gordon03}, widely used for high-redshift galaxies and reddened AGN \citep{capak15,reddy15,reddy18,hopkins04,kokorev23c,taylor25}. Assuming an intrinsic ratio of (8448/11290)$_{\rm int} = 1.336$ \citep{osterbrock89}, we infer $A_V = 0.1 \pm 0.3$ mag, consistent with negligible extinction. Despite this, the optical continuum slope is red ($\beta{\rm opt}\sim0.35$), indicating that the 
observed SED itself is intrinsically red, rather than reddened by dust.

\begin{figure}[h]
\begin{center}
\includegraphics[width=.49\textwidth]{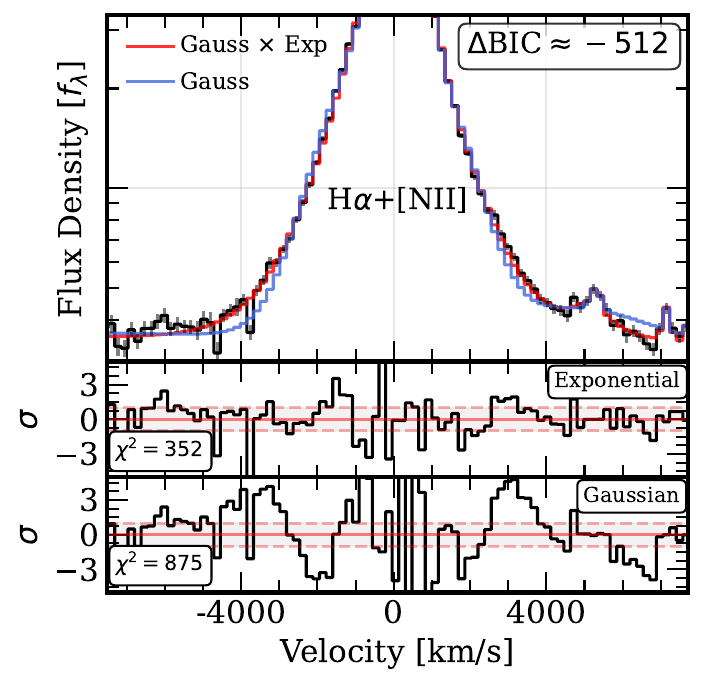}
\caption{\textbf{Exponential wings are required.} Comparison of a Gaussian$\times$Exponential model (red) and a single Gaussian profile (blue) for H$\alpha$. Uncertainties are shown with vertical lines in each spectral bin, albeit are too small to be visible. The exponential model provides a far superior fit, with smoother residuals and a strongly preferred $\Delta$BIC, highlighting the necessity of exponential wings to capture the broad-line shape.}
\label{fig:fig_exp}
\end{center}
\end{figure}

\section{Line Properties}
\label{sec:emline_props}
Our fitting procedure yields over forty emission/absorption features, most of them are detected at a high (S/N$>3$) significance.  With all the pieces in place, we now would like to comment of the line profiles, kinematics as well as the specific line species that we find.

\subsection{Exponential Wings}
\label{sec:ele}
As already noted throughout \autoref{sec:data_analysis}, and further highlighted in \autoref{fig:fig_lines}, all of the permitted lines, with the exception of O\,\textsc{i}$\lambda8448$ which is likely impacted by data quality issues, in the spectrum are better fit when the broad line is convolved with an exponential profile. The statistical preference for this (as measured by the $\Delta$BIC) being especially overwhelming for our brightest line - H$\alpha$. To further demonstrate this, and compare this fit to a more standard - Gaussian only approach, we show both of the models in \autoref{fig:fig_exp}. If the broad line were to be only fit with a Gaussian profile, large portions of the line appear to be under-fit at a $\gtrsim3$ $\sigma$ significance level, while a model that includes exponential wings shows a much smoother, albeit not perfect, residual plot. 

This strong preference for exponential wings is consistent with expectations from electron scattering in dense, ionized gas. Unlike Gaussian profiles, which would arise from Doppler broadening due to thermal and bulk motions around the SMBH, exponential profiles would only arise in environments dominated by electron scattering \citep[e.g.,][]{rusakov25,chang25}. The absence of strong broadening in the forbidden (e.g. $[$S\,\textsc{ii}$]$, $[$S\,\textsc{iii}$]$ and $[$N\,\textsc{ii}$]$) lines already hints that the densities where these electron scattered profiles originate must already exceed multiple times that the critical density of those transitions ($n_{\rm e}>10^{6}$). Combined with the fact that these broad profiles are non-Gaussian in shape, the likely volume and column densities are likely to be even higher, at $n_{\rm e}=10^{8}$ cm$^{-3}$ and $N_{\rm e}\sim10^{24}$ cm$^{-2}$, respectively, as suggested by both \citet{rusakov25} and \citet{inayoshi25}.

It is important to note that such detailed profile decomposition is only possible due to the exceptionally high S/N per pixel achieved in the line wings of our NIRSpec/G395M spectrum. As emphasized in \citet{rusakov25}, these features would be impossible to distinguish with shallower data or lower-resolution modes such as NIRSpec/PRISM or even G395M exposures lacking comparable depth. 

\begin{figure}[h]
\begin{center}
\includegraphics[width=.49\textwidth]{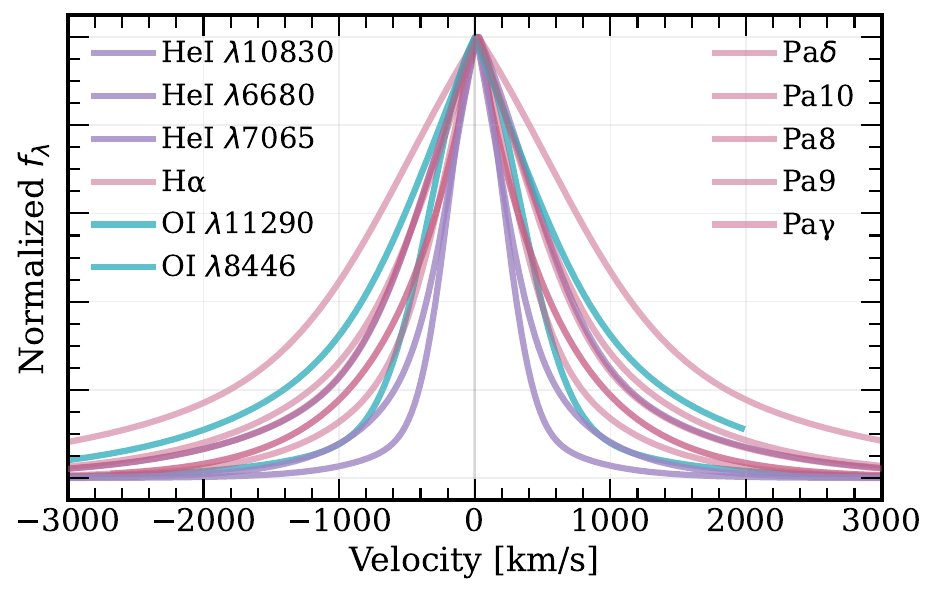}
\caption{Best-fit profiles of all broad lines, oversampled and shifted to a common center. Hydrogen recombination (H$\alpha$ and Paschen series) and O\,\textsc{i} lines show largely similar widths, consistent with their coupling through charge exchange. By contrast, He\,\textsc{i} lines are systematically narrower, likely reflecting their distinct metastable triplet physics and formation in a less dense, outer region.}
\label{fig:fig_prof}
\end{center}
\end{figure}

\subsection{Line Profiles}
Further, in \autoref{fig:fig_prof} we compare the best-fit models for all broad lines. As noted previously, with the exception of O\,\textsc{i}$\,{\lambda8448}$, every broad line is better described by exponential wings, resulting in very similar overall line shapes. The main differences arise in their widths. Despite some fits having sizable uncertainties, a trend emerges across line species. Hydrogen recombination lines (with the exception of Pa$\delta$) and O\,\textsc{i} transitions consistently show comparable FWHM values of $\sim$800 km s$^{-1}$. This agreement is not coincidental. Neutral oxygen and ionized hydrogen have nearly identical ionization potentials, enabling rapid charge-exchange coupling between O$^{+}$ + H$^{0}$ $\rightleftharpoons$ O$^{0}$ + H$^{+}$ \citep{osterbrock06}. This continual electron exchange tightly locks the spatial distribution, ionization state, and kinematics of neutral oxygen to ionized hydrogen (and vice versa), so their broad-line profiles are naturally expected to track one another. The observed similarity therefore provides an important consistency check, reinforcing that the O\,\textsc{i} emission originates in the same dense, ionized gas as the hydrogen recombination lines. In contrast, the Fe\,\textsc{ii} lines exhibit significantly narrower widths, matching the narrow cores of the permitted transitions rather than their exponential wings. This implies that the Fe\,\textsc{ii} emission arises in a cooler, less turbulent zone exterior to the scattering-dominated region, but still closely coupled to the AGN continuum source.

By contrast, He\,\textsc{i} emission is systematically narrower. He\,\textsc{i} lines such as $\lambda7065$ and $\lambda10830$ do not couple to hydrogen via charge exchange, and their lower metastable level (2$^3$S) has distinct population physics, being fed by both collisional and recombination pathways. Further, one has to take into account the radiative transfer effects, which make it possible to radiatively excite meta-stable ground state electrons. As a result, He\,\textsc{i} emission may arise from a somewhat different spatial or kinematic region than H and O. We return to this point in more detail below.

Taken together, the similar FWHM of hydrogen and oxygen broad-line profiles, combined with the near-universal exponential wings seen across all permitted transitions, strongly suggests that the line shapes are set by a common, line‐independent scattering kernel rather than by transition‐specific processes. Electron (Thomson) scattering in a dense, ionized cocoon provides a natural explanation: it produces exponential wings of nearly identical form across species, with widths set primarily by the electron temperature and column density (typically $\sim10^{24}$~cm$^{-2}$), and leaves only secondary variations from species-specific excitation or optical depth effects \citep[e.g.][]{chang25,rusakov25}. In this framework, the narrower He\,\textsc{i} lines reflect stratification within the cocoon, where He\,\textsc{i} emission arises from an outer region that has a lower density temperature or column density portion of the object, 
while the bulk hydrogen and oxygen emission share a common kinematic imprint within an inner - denser region.

\begin{figure}[h]
\begin{center}
\includegraphics[width=.49\textwidth]{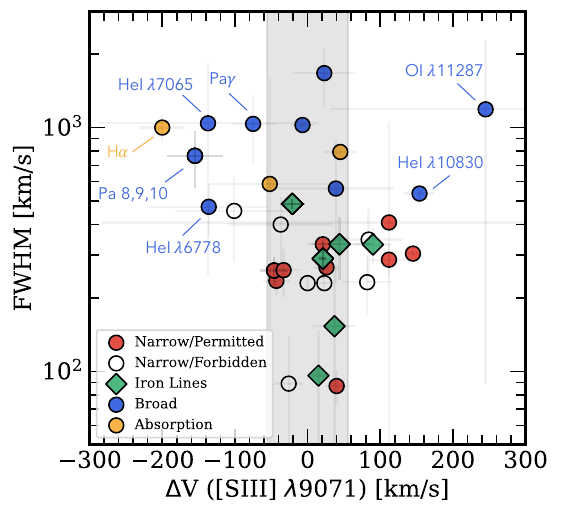}
\caption{Velocity offsets (relative to [S\,\textsc{iii}]${\lambda9071}$) and FWHM for all detected lines. Narrow permitted lines are shown as maroon circles (open for forbidden), broad lines in blue, absorption in gold, and Fe\,\textsc{ii} lines as green diamonds. The shaded band marks the velocity uncertainty set by the median spectral resolution. Narrow lines align closely with the systemic redshift, while absorption features show moderate blueshifts of $\sim150$ km s$^{-1}$.}
\label{fig:fig_offset}
\end{center}
\end{figure}

\subsection{Line Kinematics}
\label{sec:linekin}
Before we proceed to discussing individual features in more detail, we would like to comment on the systematic velocity offsets between various line species in the spectrum on \srcname. As we mentioned before, we choose the redshift of the narrow forbidden [S\,\textsc{iii}]$\,{\lambda9701}$ as the systemic one, due to it being relatively bright and isolated. We place all velocity offsets in context, by plotting all values from \autoref{tab:kinematics_narrow} and \autoref{tab:kinematics_broad} in \autoref{fig:fig_offset}.

While every fit formally yields a centroid, the significance of any offset is constrained by the spectral resolution and S/N of the data. To gauge which velocity shifts are meaningful, we adopt a conservative uncertainty of $\Delta V \sim \mathrm{med}(R)/5$, with $R = \lambda/\Delta\lambda$. This corresponds to $\sim$60 km s$^{-1}$, below which offsets are unlikely to be statistically significant.

Qualitatively, we do not find systematic shifts of either permitted or forbidden narrow lines relative to the systemic redshift. Iron lines also appear consistent with the systemic velocity. In contrast, the permitted broad lines, together with their associated absorption components, show a consistent blueshift of order $\sim$100 km s$^{-1}$.

Such blueshifts are commonly interpreted as signatures of outflowing gas in the broad-line region, where scattering and absorption occur preferentially along outflowing sight-lines. In the context of \srcname, this modest but systematic offset strengthens the case for a stratified cocoon, while the narrow lines trace gas in the outer regions at systemic velocity, the broad-line region and absorbing layers appear to be participating in bulk outflows. Confirming the detailed velocity structure, however, will require higher-resolution follow-up with the NIRSpec H-gratings \citep[e.g. see][]{saldanalopez25,torralba25}.

\subsection{Hydrogen Absorption Lines}
\label{sec:abs}
Although no strong hydrogen absorption lines are explicitly detected, the H$\alpha$ profile (\autoref{fig:fig_lines}) shows a noticeable asymmetry, with suppressed flux on the blue side of the line. This feature is best modeled by a blueshifted ($\sim-200$ km s$^{-1}$) absorption component, as already noted in \autoref{sec:data_analysis}. Similar non-resonant absorption signatures (e.g., Balmer lines) are now routinely observed in LRDs across a wide range of redshifts \citep[e.g.][]{greene24,lin24,labbe24,taylor25,kokorev24b}. The presence of such absorption implies high gas densities of order $n \sim 10^9$ cm$^{-3}$ \citep{inayoshi25,hall07}. Interestingly, only atoms in the $n=2$ state appear significantly populated, as we detect no evidence for Paschen absorption in any of the five lines present in our spectrum. This is also not an instrumental effect of the M-grating, as we clearly see an absorption feature in He\,\textsc{i}\,$\lambda10830$, but not the immediately adjacent Pa$\gamma$. This indicates that $n=3$ state is comparatively underpopulated. 

Another key manifestation of the same physical mechanism is the Balmer break, corresponding to the $n=\infty \rightarrow n=2$ transition limit. A prominent break is common in many LRD spectra \citep{setton24}, though not ubiquitous \citep[e.g.][]{kokorev23c,tripodi25}. While our spectral coverage does not extend to the Balmer limit itself, we observe a discontinuity of $\sim$1 mag between the F200W and F150W filters. Due to the F200W coverage of the SED, this  flux jump cannot be attributed to line boosting from H$\beta$ and the [O\,\textsc{iii}] doublet. Adopting the break parametrization of \citet{naidu25_bh*} we use our best-fit SED and  measure $f_{\nu,4050\textrm{\AA{}}}/f_{\nu,3670\textrm{\AA{}}}=2.02\pm0.10$. Although weaker than the extreme values ($\sim4-7$) reported for LRDz9 \citep{taylor25}, MoM BH$^*$-1 \citep{naidu25_bh*}, or ``the Cliff'' \citep{degraaff25_cliff}, this value lies close to the maximum achievable by evolved stellar populations \citep{wang24,labbe24}. 

Taken together, the presence of asymmetry in the H$\alpha$ profile, likely caused by blueshifted Balmer absorption, and a Balmer break adds to the evidence, complementing the exponential line wings, that the emission in \srcname\ arises from a dense, ionized cocoon of gas. The lack of a detectable Paschen break is consistent with this picture, as high densities and optical depths result in Thomson scattering, but would also naturally wash out higher-order continuum edges \citep[e.g.][]{wang25}. Assuming LTE, an extremely low ratio $n_{3}/n_{2}\lesssim0.01$ corresponds to an electron temperature of $T_{\rm e}\sim5000$ K or below, consistent with the warm, partially ionized gas expected in dense LRD cocoons.

We also observe moderately prominent and extremely prominent blueshifted absorption, respectively, in He\,\textsc{i}$\,{\lambda7065}$ and He\,\textsc{i}$\,{\lambda10830}$ the mechanism for this is similar, but not identical to the hydrogen lines. We also note the prominent blueshifted absorption seen in both He\,\textsc{i}$\,{\lambda7065}$ and He\,\textsc{i}$\,{\lambda10830}$. However, the physical mechanism driving these helium features differs fundamentally from that of the hydrogen lines, and we return to this in the following section.

\subsection{Helium Lines}
\label{sec:hei}
Both He\,\textsc{i}$\,{\lambda7065}$ and He\,\textsc{i}$\,{\lambda10830}$ show blueshifted ($\sim$70–80 km s$^{-1}$ from the narrow-line center) absorption components with widths of $\sim$600–700 km s$^{-1}$. Unlike the Balmer series, helium transitions are resonantly scattered, so the arguments about $n=2$ state abundances do not apply. Before addressing the absorption itself, it is useful to consider what the observed line strengths already tell us.

In the spectrum of \srcname\ we detect three He\,\textsc{i} lines: ${\lambda6680}$, ${\lambda7065}$, and ${\lambda10830}$. These originate from different physical mechanisms. The ${\lambda6680}$ singlet line is produced via recombination or radiative pumping and is only weakly dependent on density. By contrast, ${\lambda7065}$ and ${\lambda10830}$ belong to the triplet system (along with ${\lambda3889}$, not covered by G395M), where the lower level is populated due to its very long lifetime. Electrons in this ``ground state'' can be 
can be collisionally or radiatively excited into higher energies and therefore the triplet lines strengths depend strongly on density and, to a lesser extent, temperature. As emphasized by \citet{berg25}, He\,\textsc{i}$\,{\lambda10830}$ is the most sensitive density diagnostic among these transitions, followed by ${\lambda7065}$. The fact that both are much stronger than ${\lambda6680}$ indicates that the region where they form is very dense. The prominence of ${\lambda10830}$ in particular already points to a high-density environment with a temperature $T\gtrsim10^4$ K. Further, the broad ${\lambda6680}$ has a noticeably higher FWHM compared to the triplet lines, but is consistent with hydrogen recombination lines. Which, again, makes sense since both originate from the same mechanism.

So what about absorption? Well, it turns out that its presence further reinforces this picture. The lower level (2$^3$S) of the triplet system is a metastable, long-lived state that effectively acts as a ground state in dense or partially ionized gas. As a result, photons from the 2$^3$P $\rightarrow$ 2$^3$S transition can be resonantly absorbed and re-emitted many times, analogous to Ly$\alpha$ or Mg\,\textsc{ii}. This mechanism naturally produces strong absorption features when the He\,\textsc{i} column density is high. The fact that He\,\textsc{i}$\,{\lambda10830}$ not only dominates the helium spectrum in emission but also shows the deepest absorption in the entire spectrum provides compelling evidence for a dense, ionized cocoon of gas enshrouding \srcname.

\begin{figure*}
\begin{center}
\includegraphics[width=.95\textwidth]{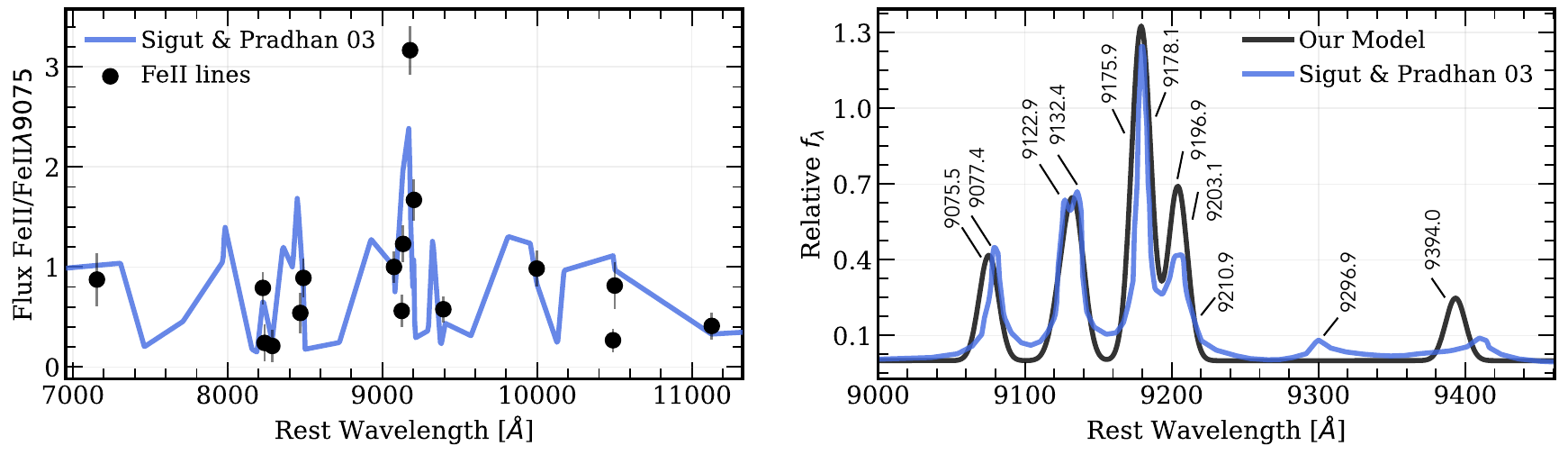}
\caption{\textbf{Ly$\alpha$–pumped iron emission in \srcname.}
\textbf{Left:} Observed Fe\,\textsc{ii} flux ratios (black points), normalized to Fe\,\textsc{ii} $\lambda9075$, compared with predictions from Ly$\alpha$ fluorescence models of \citet{sigut03} (blue). The close correspondence across $\sim14$ lines indicates a common excitation mechanism driven by Ly$\alpha$ pumping in dense, partially shielded gas.
\textbf{Right:} Zoom-in on the $\lambda\lambda 9000–9400$ \AA\ complex showing the remarkable agreement between the modeled continuum-subtracted Fe\,\textsc{ii} spectrum (black) and the \citet{sigut03} model (blue). We show air wavelengths of each Fe\,\textsc{ii} line directly form  \citeauthor{sigut03}, including the blended features. Together, these comparisons support an origin in the dense inner regions of the cocoon surrounding the accreting black hole.}
\label{fig:fig_iron}
\end{center}
\end{figure*}

\subsection{Oxygen Lines}
\label{sec:oi}
Above, we noted that O\,\textsc{i}$\,{\lambda8446}$ and ${\lambda11290}$ share very similar profiles with the hydrogen recombination lines. This is naturally expected because neutral oxygen is tightly coupled to hydrogen via charge exchange, which rapidly equilibrates the O/H ionization states in dense gas. When charge exchange is fast (high $n_{\rm H}$), the O\,\textsc{i}/O\,\textsc{ii} ratio tracks the H\,\textsc{i}/H\,\textsc{ii} ratio, so the O\,\textsc{i}–emitting gas shares the same kinematics as the hydrogen recombination region \citep{draine11}. The observed agreement in FWHM  between H and O\,\textsc{i} lines is therefore an important consistency check that both species originate in the same dense, partially ionized phase.

A second, independent clue of high gas density and radiation field intensity comes from the excitation mechanism. The O\,\textsc{i} $\lambda8446$–$\lambda11287$ pair is the classic signature of Ly$\beta$ (Bowen) fluorescence: Ly$\beta$ pumps O\,\textsc{i} from the ground state and the ensuing cascade preferentially populates the levels that emit 1.129 $\mu$m that then cascade down through the 8446~\AA\ line. In this channel, the line emissivity scales with both the neutral oxygen (or equivalently neutral hydrogen, via charge exchange) column and the local Ly$\beta$ radiation field, so producing strong O\,\textsc{i} fluorescence requires both a very bright Ly$\beta$ source and a dense neutral (or partially ionized) cocoon.

Further, we do not detect ($5\sigma$ upper limit $< 12/  [10^{-20}$ erg s$^{-1}$cm$^{-2}$]) any other of the permitted O\,\textsc{i} $\lambda7774$, $\lambda7254$, and $\lambda7790$ lines, and the [O\,\textsc{i}]\,$\lambda6302$/O\,\textsc{i}\,$\lambda8446$ ratio is weak. If collisional excitation or pure recombination dominated, these lines would be comparatively strong; their absence strongly favors Bowen fluorescence as the primary driver of the observed O\,\textsc{i} emission.

Finally, the velocity structure adds a natural stratification: hydrogen recombination and O\,\textsc{i} (fluorescent, charge-exchange coupled) share widths and profiles, while He\,\textsc{i} lines are systematically narrower. Since He\,\textsc{i} triplet transitions emerge from a metastable 2$^3$S level (with distinct, density-sensitive population pathways and resonant transfer), they likely trace a kinematically distinct layer within the same cocoon. 

\subsection{Ly$\alpha$ Fluorescence in Iron Lines}
\label{sec:fe}
Permitted iron emission, primarily Fe\,\textsc{ii} and Fe\,\textsc{iii} \citep[e.g.][]{labbe24,tripodi25,torralba25}, but in some cases extending to highly ionized species such as Fe\,\textsc{vii} \citep{tang25,lambrides25} and Fe\,\textsc{x} \citep{furtak24}, has become a recurring feature in LRD spectra. In \srcname, the spectrum is exceptionally rich in near-infrared iron lines: we identify 16 Fe\,\textsc{ii} and one Fe\,\textsc{iii} transition. Understanding the origin of this emission is key to interpreting the dense gas environment in LRDs.

The physics of Fe\,\textsc{ii} emission has long been a challenge for BLR photoionization models \citep{joly93}. Thick, high-column density ($\sim10^{25}$ cm$^{-2}$) gas at the edges of the accretion disk has been invoked as a potential source \citep{joly87,collin-souffrin88}, where the scattering and absorption of the hard X-ray photons would ionize the gas. Similarly, this would also enhance the Balmer and Paschen line luminosities. Further, extensive theoretical work has shown that Ly$\alpha$ fluorescence is fundamental in reproducing the observed Fe\,\textsc{ii} strengths \citep{sigut98,sigut03}. Ly$\alpha$ pumping not only boosts the UV and optical Fe\,\textsc{ii} emission, but also predicts strong lines in the near-infrared.

This expectation aligns closely with our observations. In \autoref{fig:fig_iron}, we compare our measured Fe\,\textsc{ii} flux ratios (normalized to Fe\,\textsc{ii}\,$\lambda9075$) against the Ly$\alpha$-pumped model predictions of \citet{sigut03}. With the exception of Fe\,\textsc{ii}\,$\lambda9179$, affected by blending with Fe\,\textsc{ii}\,$\lambda9204$ and Pa9, we find remarkably good agreement across the suite of detected features. In particular, the dense forest of Fe\,\textsc{ii} transitions spanning $\lambda\lambda9000$–9200 \AA\ (center-right panel in \autoref{fig:fig_lines}) is reproduced almost exactly by the theoretical spectrum \citep[cf.\ Fig. 11 in][]{sigut03}. We overlay the \citeauthor{sigut03} model directly on our continuum-subtracted fit in \autoref{fig:fig_iron}, demonstrating that the observed Fe\,\textsc{ii} emission is fully consistent with fluorescence in dense gas near the broad-line region. The implication is that the majority of the Fe\,\textsc{ii} emission in \srcname\ arises through Ly$\alpha$ pumping, directly tracing an extremely dense medium, ionized by an extremely luminous source - likely an accreting black hole. 
Given that the Fe\,\textsc{ii} are somewhat narrower compared to other permitted features, it is likely that the emitting region is further from the BLR, something that has already been shown in a classic example of narrow Fe\,\textsc{ii} emitter - I Zw 1 \citep{rudy00,marinello16}. 

Together with the Balmer absorption, helium triplet features, and Bowen fluorescence in O\,\textsc{i}, the iron forest adds another piece of evidence that points to a dense cocoon enshrouding the black hole.

\subsection{Source Properties}
\label{sec:bh_prop}
Based on the previous discussion, we now present the likely physical properties of the black hole and its host. The coexistence of broad permitted and narrow forbidden lines indicates, as in most LRDs \citep{hviding25}, that the broad components originate in an AGN broad-line region (BLR). Following \citet{greene05}, we derive $M_{\rm BH}$ from the luminosity and width of the broad H$\alpha$ line, assuming the exponential-wing profile demonstrated in \autoref{fig:fig_lines} and \autoref{fig:fig_exp}. With FWHM${\rm H\alpha}=1024\pm21$ km s$^{-1}$, negligible dust, and $\mu=2.04$, we obtain $\log(M_{\rm BH}/M_\odot)=6.65\pm0.15$, where the dominant uncertainty arises from the \citeauthor{greene05} calibration. A purely Gaussian fit, though statistically disfavored, would yield a FWHM three times larger and a black hole mass nearly a dex higher. Throughout this work we adopt the $M_{\rm BH}$ from the exponential model.

Further, by assuming that the bolometric luminosity ($L_{\rm bol}$) scales as $L_{\rm bol} = 130 \times L_{\rm H\alpha}$ \citep{richards06}, we find $L_{\rm bol}=(1.06\pm0.14)\times10^{45}$ erg/s. Using our $M_{\rm BH}$ derived by assuming exponential wings, we then find that this object is accreting at a super-Eddington rate - $L_{\rm bol}/L_{\rm edd}=1.86\pm0.25$. This is higher than the vast majority of LRDs at all redshifts \citep[e.g.][]{taylor25,kokorev23c,juodzbalis24_rosetta,akins25,furtak24,maiolino23b,kocevski24}, which are found to accrete at sub-Eddington rates. However recent works examining LRDs with extreme Balmer breaks derive accretion rates that exceed the Eddington limit \citep{degraaff25_cliff,naidu25_bh*,lambrides24}.

A recent work examining multi-wavelength LRD data however has suggested that the bolometric luminosity emerges mostly from the rest-frame optical, with X-rays and radio contributions being largely subdominant \citep{greene25}. Given the suggested dominance of the optical light in LRDs, the much lower bolometric corrections ($\times7-10$ lower) would potentially imply lower black hole masses and total bolometric luminosities (e.g. derived from H$\alpha$) than the standard \citep[e.g][]{greene05} might suggest. This adjustment would lower both the inferred black hole masses and bolometric luminosities (e.g., those derived from H$\alpha$). However, because the two quantities scale together, the implied super-Eddington nature of \srcname\ would remain unchanged, though the black hole mass could decrease to $\sim10^{5.5}$–$10^{5.8}\,M_\odot$. A more detailed reassessment of bolometric corrections for LRDs is clearly warranted, but such an analysis lies beyond the scope of this work. For consistency with previous studies, we adopt the standard AGN bolometric corrections throughout, with all derived parameters listed in \autoref{tab:tab1}.

Finally, we estimate an upper limit on the stellar mass in \srcname. Previous studies have modeled LRDs using joint galaxy+AGN SED decomposition \citep[e.g.,][]{furtak24,kokorev23c}, and more recently within the BH$^*$ framework have assumed that most of the rest–UV flux arises from the host galaxy \citep{taylor25,naidu25_bh*}. Our spatial decomposition (\autoref{fig:fig_morph}) supports this assumption: the extended component contributes $>80$\% of the rest–UV light, implying that the bulk of the stellar mass resides in this resolved structure. Given this, we adopt a simple empirical approach using the $M_{\rm UV}$–$M_*$ relation \citep[e.g.][]{stark09,labbe13} and obtain a conservative upper limit on the stellar mass of $M_* \lesssim 10^{7.5} M_\odot$. This in return gives us a BH-to-host mass ratio of $\lesssim 0.14$, which is significantly elevated from local expectations \citep{greene05}, but is not as extreme as some other LRDs reported in the literature \citep{furtak24,kokorev23c,maiolino23b}.

\begin{figure*}
\begin{center}
\includegraphics[width=.95\textwidth]{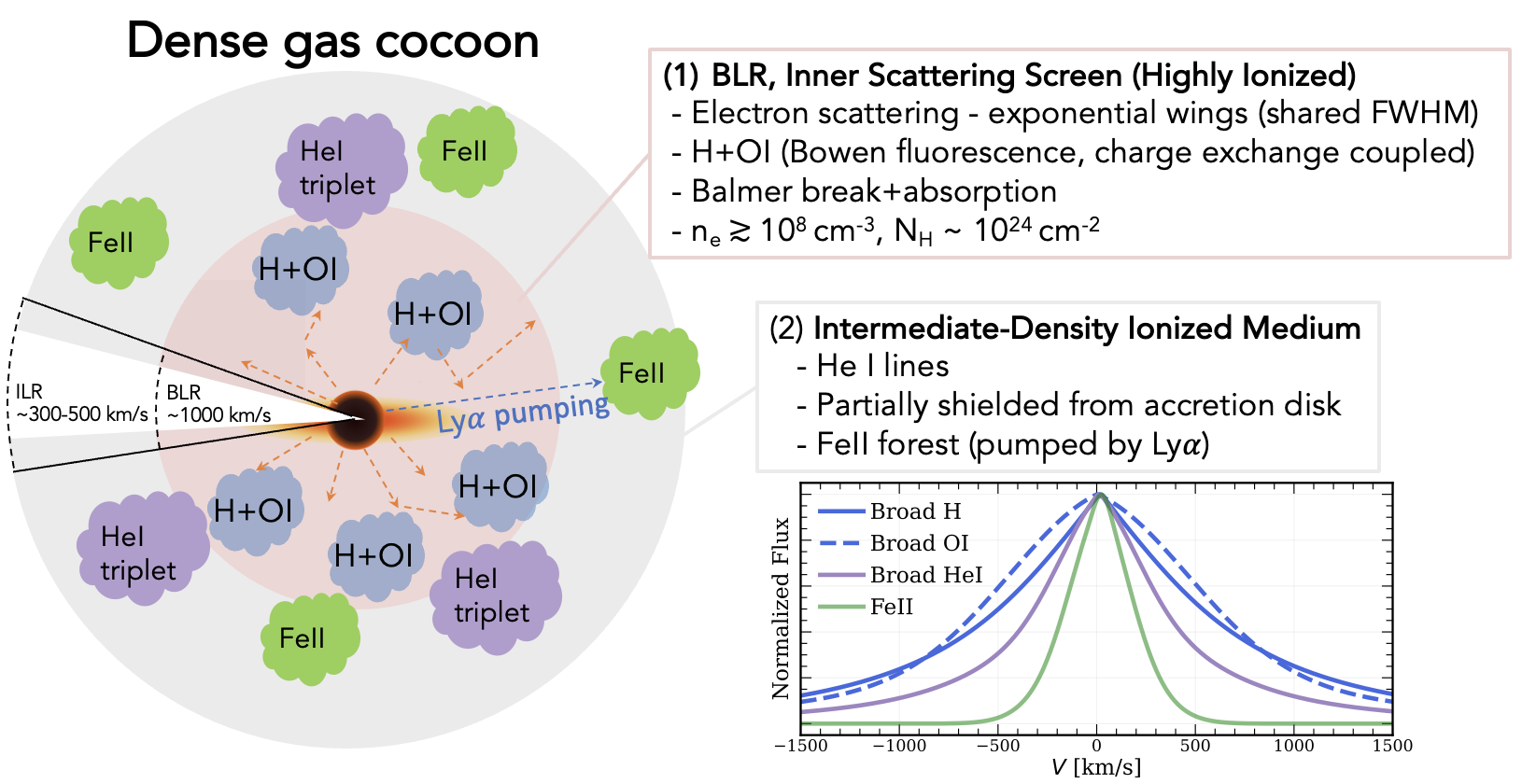}
\caption{\textbf{Physical picture of the dense cocoon around \srcname.}
The inner region (1) corresponds to the dense, highly ionized broad-line region where electron scattering (orange lines) produces exponential wings, H and O I share coupled kinematics (via charge exchange and Bowen fluorescence), and strong Balmer absorption/break signatures arise. The outer layer (2) represents an intermediate-density ionized medium, where He I triplet lines show resonant absorption from the metastable 2$^3$S level and Fe II emission is driven by Ly$\alpha$ fluorescence. Together, these zones form a stratified cocoon enshrouding the accreting black hole. Bottom right panel shows stacks line profiles of H, O\,\textsc{i}, He\,\textsc{i} and Fe\,\textsc{ii}.}
\label{fig:fig_cartoon}
\end{center}
\end{figure*}

\section{Discussion and Summary} \label{sec:concl}

\subsection{Dense and Ionized Gas Surrounding the AGN}
\label{sec:dgas}

Our NIRSpec/G395M observations of \srcname\ reveal a remarkably consistent picture: across independent tracers, the line emission requires an environment of extremely high density and partial ionization.

First, the broad permitted lines are universally better fit by exponential wings, a hallmark of electron scattering in gas with $n_{\rm e}\gtrsim10^8$ cm$^{-3}$ and column densities approaching $N_{\rm e}\sim10^{24}$ cm$^{-2}$. Such profiles are not reproduced by Doppler broadening alone and point to an ionized scattering medium enveloping the source. The fact that hydrogen and O\,\textsc{i} lines share consistent FWHM values further anchors this interpretation, as charge exchange tightly couples neutral oxygen to the ionization and kinematics of hydrogen.

Second, the detection of blueshifted Balmer absorption and a significant Balmer break both require high $n=2$ populations and densities $n\sim10^9$ cm$^{-3}$. Helium transitions provide a complementary view: the triplet lines $\lambda7065$ and $\lambda10830$ are both strongly enhanced relative to the singlet states and show deep blueshifted absorption, consistent with resonant scattering from the metastable 2$^3$S level. Their systematically narrower widths compared to hydrogen and oxygen suggest stratification, with helium arising from denser, more compact layers of the cocoon.

Third, the O\,\textsc{i}$\,\lambda8446$–$\lambda11290$ pair confirms Ly$\beta$ fluorescence, requiring both a bright Ly$\beta$ radiation field and a dense reservoir of neutral gas. The absence of other permitted O\,\textsc{i} lines not fed by Ly$\beta$ strengthens this conclusion. Finally, the detection of 16 Fe\,\textsc{ii} lines, forming an incredibly rich iron forest in this LRD, matches predictions from Ly$\alpha$ fluorescence models, again demanding an intense radiation field and very high densities.

Because the BLR is unresolved in essentially all AGN, broad-line widths are traditionally interpreted as virial tracers of the black hole potential. In \srcname, the virial story alone is insufficient: the issue is not how broad the lines are, but how they broaden. The profiles exhibit extended, nearly linear wings in velocity space that are incompatible with a Gaussian. Instead, the lines are systematically and significantly better described by a model consisting of a narrow Gaussian core (virial motion) plus exponential wings, the hallmark of Thomson scattering in a dense ionized medium. Thus, the line shape encodes both gravitational kinematics and radiative-transfer physics in the surrounding cocoon. Our schematic (\autoref{fig:fig_cartoon}) summarizes this revised view: in LRDs, broad-line profiles are not set by dynamics alone, but by the scattering environment through which the photons escape.

Such conditions—high optical depths, large column densities, and evidence for radiation-dominated gas—are precisely those expected in super-Eddington accretion flows. In this regime, radiation pressure inflates the inner accretion structure, driving powerful winds and forming the very dense, partially ionized envelope we infer here. The low X-ray luminosities and weak radio emission commonly observed in LRDs \citep[e.g.,][]{akins24,ananna24,kokubo24,yue24} are consistent with this picture: the X-rays are likely absorbed or thermalized within the optically thick cocoon, while dust cannot survive in such an intense radiation field \citep[e.g.,][]{inayoshi25,naidu25_bh*,degraaff25_cliff,taylor25,lambrides24}. Thus, the spectroscopic signatures observed in \srcname—exponential wings, absorption features, and fluorescence-driven metal lines—fit naturally into a scenario where super-Eddington accretion onto a low-mass black hole powers the luminous yet heavily reprocessed emission.

Taken together, the exponential wings, Balmer and helium absorption, Bowen oxygen lines, and Ly$\alpha$-pumped iron forest all converge on the same physical picture: \srcname\ is enshrouded in a dense, partially ionized cocoon of gas.

\subsection{Final Remarks}
\label{sec:remarks}
Using a combination of the intrinsically deepest NIRCam photometry and NIRSpec/G395M observations of the lensed AS1063 field, we present a detailed investigation of a little red dot at $z=3.501$. The spectrum of \srcname\ is exceptionally rich, with more than forty detected features, allowing us to probe the physical conditions of the gas with unprecedented detail. Multiple independent diagnostics converge on the presence of a dense, partially ionized cocoon heated by a powerful ionizing source.

Typical of other LRDs, \srcname\ is extremely compact in the rest-optical ($r_{\rm eff}<300$ pc) but exhibits more extended structure in the rest-UV ($r_{\rm eff}\sim1000$ pc; \autoref{fig:fig1}). It shows unmistakable AGN signatures through broad permitted lines \citep[e.g.,][]{greene24,kocevski23,kokorev23c,furtak24}. The uniquely deep G395M spectrum further reveals clues to the physical origin of both its continuum and line-emission properties. We detect exponential broad-line wings, Balmer and helium absorption, Bowen--pumped O\,\textsc{i}, Fe\,\textsc{ii} emission produced by Ly$\alpha$ fluorescence, and evidence for rapid, potentially super-Eddington growth.

Together, these diagnostics provide one of the clearest cases yet for the BH$^*$ ``dense cocoon'' scenario \citep{inayoshi25,naidu25_bh*,degraaff25_cliff,taylor25}. The same ingredients we observe---exponential broad wings, Balmer-break absorption, He\,\textsc{i} absorption, and rich Fe\,\textsc{ii} and O\,\textsc{i} fluorescence---have been detected individually in several other bright LRDs with medium- and high-resolution spectroscopy \citep[e.g.][]{torralba25,deugenio25,labbe24,furtak24,wang25,juodzbalis24_rosetta,lambrides25}, suggesting that dense, optically thick gas may be a common feature of the population rather than an anomaly. What distinguishes \srcname\ is that all signatures are captured simultaneously and at high S/N, allowing a self-consistent physical interpretation. If such cocoons are widespread, then super-Eddington accretion may be a typical pathway for black-hole growth in LRDs, especially among the most luminous systems. Establishing how these signatures vary with luminosity and redshift will be essential for determining whether dense cocoons represent a dominant mode of early black-hole assembly.

The convergence of five independent diagnostics---exponential scattering wings, Balmer-limit absorption, helium triplet physics, and two distinct fluorescence channels---leaves little doubt: \srcname\ hosts a dense ($n\sim10^{8-9}$ cm$^{-3}$), optically thick ($N_e\sim10^{24}$ cm$^{-2}$) cocoon of partially ionized gas surrounding a super-Eddington accreting black hole. This represents some of the most direct and comprehensive spectroscopic evidence to date for the dense cocoon scenario in LRDs.

\acknowledgements
The authors would like to thank Aaron Sigut and Anil Pradhan for their help with understanding iron emission in AGN. The authors would like to acknowledge the National Institute of Standards and Technology (NIST) database of spectral lines \citep{nist}, which made identification of less-known emission features possible. VK, JC, SF, DB, LF, TH and JM acknowledge support from the University of Texas at Austin Cosmic Frontier Center. AZ acknowledges support by the Israel Science Foundation Grant No. 864/23.  This work is based on observations made with the NASA/ESA/CSA \textit{James Webb Space Telescope}, obtained at the Space Telescope Science Institute, which is operated by the Association of Universities for Research in Astronomy, Incorporated, under NASA contract NAS5-03127. The JWST data presented in this article were obtained from the Mikulski Archive for Space Telescopes (MAST) at the Space Telescope Science Institute. The specific observations analyzed can be accessed via \dataset[doi: 10.17909/4byn-fe55]{https://doi.org/10.17909/4byn-fe55} and \dataset[doi: 10.17909/zq0c-8t87]{http://dx.doi.org/10.17909/zq0c-8t87}. These observations are associated with programs GO \#3293 and DDT \#9223. 

\software{EAZY \citep{brammer08}, grizli \citep{grizli}, msaexp \citep{msaexp}, photutils \citep{photutils}, pysersic \citep{pasha23}, sep \citep{sep}, SExtractor \citep{sextractor}} 

\facilities{\jwst, \hst}

\clearpage


\bibliographystyle{aasjournal}
\bibliography{refs}

\end{document}